\PassOptionsToPackage{table}{xcolor} % Add options here (e.g., table, svgnames, dvipsnames)

\documentclass[webpdf,contemporary,large]{oup-authoring-template}%

\usepackage{xcolor}

\renewcommand{\paragraph}[1]{\smallskip\noindent{\bf #1}}
\usepackage{multirow}
\usepackage{natbib}
\usepackage{wrapfig}

\DeclareMathAlphabet\mathbfcal{OMS}{cmsy}{b}{n} %bold mathcal

\usepackage{stackrel,amssymb,amsmath}
\usepackage{graphicx}
\usepackage[margin=1in]{geometry}
\usepackage{array}

\definecolor{darkgreen}{RGB}{34, 139, 34} 
\definecolor{ipcolor}{RGB}{242,170,60}
\definecolor{bpcolor}{RGB}{55,126,247}
\usepackage{xspace}
\usepackage{mathtools}
\usepackage{hyperref}  % citation link
\hypersetup{
    colorlinks=true,
    linkcolor=blue, % color of internal links
    urlcolor=blue,  % color of external links
    citecolor=blue, % color of citation links
}

\renewcommand{\cite}[1]{\citep{{#1}}}  

\usepackage{verbatim}

\usepackage{algorithm}
\usepackage{algpseudocode}

\usepackage{courier}
\usepackage{float}
\usepackage{tikz}
\usetikzlibrary{positioning, arrows.meta, calc}
\usetikzlibrary{shapes,snakes}
\usepackage[capitalise]{cleveref}
\usepackage{tablefootnote}
\usepackage{siunitx}
\usepackage{booktabs}
\usepackage{bm}

\usepackage{xspace} % ensure loaded somewhere
\usepackage{amsmath,amssymb,bm} % for \boldsymbol / \mathbb etc (ensure in main)

\renewcommand{\vec}[1]{\ensuremath{\boldsymbol{{#1}}}\xspace}

\newcommand{\vecp}{\ensuremath{\vec{p}}\xspace} % protein sequence
\newcommand{\vecx}{\ensuremath{\vec{x}}\xspace} % RNA / mRNA sequence
\newcommand{\vecy}{\ensuremath{\vec{y}}\xspace} % structure

\newcommand{\ptoken}[1]{\ensuremath{p_{#1}}\xspace}
\newcommand{\xtoken}[1]{\ensuremath{x_{#1}}\xspace}

\newcommand{\protlen}{\ensuremath{m}\xspace}        % protein length
\newcommand{\rnalen}{\ensuremath{N}\xspace}         % nucleotide length
\newcommand{\posindex}{\ensuremath{n}\xspace}

\newcommand{\defeq}{\ensuremath{\stackrel{\Delta}{=}}\xspace}

\newcommand{\N}{\ensuremath{\mathcal{N}}\xspace}
\newcommand{\nucA}{\ensuremath{\texttt{A}}\xspace}
\newcommand{\nucC}{\ensuremath{\texttt{C}}\xspace}
\newcommand{\nucG}{\ensuremath{\texttt{G}}\xspace}
\newcommand{\nucU}{\ensuremath{\texttt{U}}\xspace}
\newcommand{\Nset}{\ensuremath{\{\nucA,\nucC,\nucG,\nucU\}}\xspace}

\newcommand{\designspace}{\ensuremath{\mathcal{X}}\xspace}
\newcommand{\designspaceof}[1]{\ensuremath{\designspace(#1)}\xspace}
\newcommand{\StructSet}[1]{\ensuremath{\mathcal{Y}(#1)}\xspace}  % feasible structures

\newcommand{\bpp}[2]{\ensuremath{p_{#1,#2}}\xspace}
\newcommand{\unp}[1]{\ensuremath{q_{#1}}\xspace}

\newcommand{\one}{\ensuremath{\mathbf{1}}\xspace}
\newcommand{\ind}[1]{\ensuremath{\one\!\left[#1\right]}\xspace}

\DeclareMathOperator{\E}{\mathbb{E}}
\DeclareMathOperator{\argmin}{argmin}

\newcommand{\metricindex}{\ensuremath{k}\xspace}      % metric index
\newcommand{\numMetrics}{\ensuremath{K}\xspace}       % number of metrics

\newcommand{\rawmetrick}[1]{\ensuremath{g_{#1}}\xspace}

\newcommand{\metrictfk}[1]{\ensuremath{\phi_{#1}}\xspace}

\newcommand{\metrick}[1]{\ensuremath{f_{#1}}\xspace}

\newcommand{\weightk}[1]{\ensuremath{w_{#1}}\xspace}
\newcommand{\Obj}{\ensuremath{F}\xspace}

\newcommand{\vectheta}{\ensuremath{\boldsymbol{\theta}}\xspace}

\newcommand{\translate}{\ensuremath{\mathrm{translate}}\xspace}

\newcommand{\optx}{\ensuremath{\vecx^{\star}}\xspace}

\newcommand{\linearfold}{{LinearFold}\xspace}
\newcommand{\linearpartition}{{LinearPartition}\xspace}
\newcommand{\lineardesign}{{LinearDesign}\xspace}
\newcommand{\ensembledesign}{{EnsembleDesign}\xspace}

\newcommand{\PF}{\ensuremath{Q}\xspace}
\newcommand{\freeenergy}{\ensuremath{\Delta G^{\circ}}\xspace}

\newcommand{\GasConst}{\ensuremath{R}\xspace}
\newcommand{\Temp}{\ensuremath{T}\xspace}
\newcommand{\RT}{\ensuremath{\GasConst\Temp}\xspace}

\newcommand{\MFE}{\ensuremath{\mathrm{MFE}}\xspace}
\newcommand{\EFE}{\ensuremath{\mathrm{EFE}}\xspace}
\newcommand{\AUP}{\ensuremath{\mathrm{AUP}}\xspace}
\newcommand{\AccessU}{\ensuremath{\mathrm{AccessU}}\xspace}
\newcommand{\CAI}{\ensuremath{\mathrm{CAI}}\xspace}
\newcommand{\COMBO}{\ensuremath{\mathrm{COMBO}}\xspace}
\newcommand{\mfe}{\ensuremath{\mathrm{MFE}_\mathrm{rel}}\xspace} % normalized MFE score

\newcommand{\alphaw}{\ensuremath{\alpha}\xspace}
\newcommand{\betaw}{\ensuremath{\beta}\xspace}
\newcommand{\gammaw}{\ensuremath{\gamma}\xspace}
\newcommand{\deltaw}{\ensuremath{\delta}\xspace}

\newcommand{\latticestate}{\ensuremath{s}\xspace}
\newcommand{\outedges}[1]{\ensuremath{\mathcal{A}(#1)}\xspace}

\newcommand{\edgeact}{\ensuremath{a}\xspace}  % generic edge/action
\newcommand{\transprob}[3]{\ensuremath{p_{#1}\!\left(#2\,\middle|\,#3\right)}\xspace}

\newcommand{\seqdist}[1]{\ensuremath{p_{#1}}\xspace}

\newcommand{\logit}[2]{\ensuremath{\theta_{#1,#2}}\xspace}

\newcommand{\pathseq}{\ensuremath{\tau}\xspace}

\newcommand{\batchsize}{\ensuremath{M}\xspace}

\newcommand{\sampleindex}{\ensuremath{i}\xspace}
\newcommand{\iterindex}{\ensuremath{t}\xspace}

\newcommand{\ExpObj}{\ensuremath{J}\xspace}

\newcommand{\stepsize}{\ensuremath{\eta}\xspace}
\newcommand{\gradest}{\ensuremath{\widehat{\nabla \ExpObj}}\xspace}

\newcommand{\baseline}{\ensuremath{b}\xspace}
\newcommand{\batchstd}{\ensuremath{\sigma}\xspace}
\newcommand{\advw}{\ensuremath{A}\xspace}
\newcommand{\vareps}{\ensuremath{\varepsilon}\xspace}

\newcommand{\edgealt}{\ensuremath{u}\xspace}

\newcommand{\adamm}[1]{\ensuremath{m_{#1}}\xspace}
\newcommand{\adamv}[1]{\ensuremath{v_{#1}}\xspace}

\newcommand{\adammhat}[1]{\ensuremath{\hat m_{#1}}\xspace}
\newcommand{\adamvhat}[1]{\ensuremath{\hat v_{#1}}\xspace}

\newcommand{\adameps}{\ensuremath{\epsilon}\xspace} % or \epsilon_{\mathrm{Adam}}

\newcommand{\batchbest}{\ensuremath{\bigl(\widehat{F}_{\min}\bigr)}\xspace}

\newcommand{\bestsofar}{\ensuremath{F^*}\xspace}               % best-so-far value
\newcommand{\patience}{\ensuremath{P}\xspace}                      % patience (iters)
\newcommand{\stopminimprove}{\ensuremath{\xi_{\mathrm{stop}}}\xspace} % min improvement
\newcommand{\maxiters}{\ensuremath{T_{\max}}\xspace}               % max iterations cap
\newcommand{\objmark}{\textsuperscript{$\star$}}

\begin{document}

\journaltitle{Journal Title Here}
%\DOI{DOI HERE}
\copyrightyear{2022}
\pubyear{2019}
\access{Advance Access Publication Date: Day Month Year}
\appnotes{Paper}

\title{Sampling-based Continuous Optimization for Messenger RNA Design}

%
%\titlerunning{Abbreviated paper title}
% If the paper title is too long for the running head, you can set
% an abbreviated paper title here
%
\author[1]{\large Feipeng Yue} %\inst{1} \and
\author[1]{\large Ning Dai} %\inst{1} \and
\author[3]{\large Wei Yu Tang} %\inst{1} \and
\author[1]{\large Tianshuo Zhou} %\inst{1} \and
\author[4,5,6]{\large David Mathews} %\inst{3, 4, 5}  \and
\author[1,2,$\ast$]{\hspace{-0.2cm}\large Liang Huang} %\inst{1,2}$^{,\star}$}

\authormark{Yue et al.}
% First names are abbreviated in the running head.
% If there are more than two authors, 'et al.' is used.
%
%\address[1]{\orgdiv{School of EECS}}
%\address[2]{\orgdiv{Dept.~of Biochemistry \& Biophysics}, \orgname{Oregon State University}, \country{USA}}
%\address[3]{\orgdiv{Dept.~of Biochemistry \& Biophysics}}
%\address[4]{\orgdiv{Center for RNA Biology}}
%
%\address[5]{\orgdiv{Dept.~of Biostatistics and Computational Biology}, \orgname{University of Rochester Medical Center}, \country{USA}}

\address[1]{\orgdiv{School of EECS},
\orgname{ Oregon State University}, \orgaddress{\street{Corvallis}, \postcode{97330}, \state{OR}, \country{USA}}}
\address[2]{\orgdiv{Dept.~of Biochemistry \& Biophysics}, \orgname{Oregon State University},  \orgaddress{\street{Corvallis}, \postcode{97330}, \state{OR}, \country{USA}}}
\address[3]{\orgdiv{Dept.~of Quantitative and Computational Biology}, \orgname{University of Southern California}, \orgaddress{Los Angeles},
\postcode{90089}, \state{CA}, \country{USA}}
\address[4]{\orgdiv{Dept.~of Biochemistry \& Biophysics},
\orgname{University of Rochester Medical Center}, \orgaddress{\street{Rochester}, \postcode{14642}, \state{NY}, \country{14642}}}
\address[5]{\orgdiv{Center for RNA Biology},
\orgname{University of Rochester Medical Center}, \orgaddress{\street{Rochester}, \postcode{14642}, \state{NY}, \country{14642}}}
\address[6]{\orgdiv{Dept.~of Biostatistics \& Computational Biology}, \orgname{University of Rochester Medical Center}, \orgaddress{\street{Rochester}, \postcode{14642}, \state{NY}, \country{14642}}}

%\abstract{RNA design, the task of finding a sequence that folds into a target secondary structure, has broad biological and biomedical impact but remains computationally challenging due to the exponentially large sequence space and exponentially many competing folds. Traditional approaches treat it as an optimization problem, relying on per-instance %combinatorial 
%heuristics or constraint-based search. We instead reframe RNA design as conditional sequence generation and introduce a reusable neural approximator, instantiated as an autoregressive language model (LM), that maps target structures directly to sequences. We first train our model in a supervised setting on random-induced structure-sequence pairs, and then use reinforcement learning (RL) to optimize end-to-end metrics. We also propose methods to select a small subset for RL that greatly improves RL efficiency and quality. Across four datasets, our approach outperforms 
%state-of-the-art systems on key metrics such as Boltzmann probability while being 1.7$\times$ faster, establishing conditional LM generation as a scalable, task-agnostic alternative to per-instance optimization for RNA design.
%Our code and data are available at
%{\small \textbf{\url{https://github.com/KuNyaa/RNA-Design-LM}}}.}

\abstract{Designing messenger RNA (mRNA) sequences for a fixed target protein requires searching an exponentially large synonymous space while optimizing properties that affect stability and downstream performance. This is challenging because practical mRNA design involves multiple coupled objectives beyond classical folding criteria, and different applications prefer different trade-offs. We propose a general sampling-based continuous optimization framework, inspired by SamplingDesign, that iteratively samples candidate synonymous sequences, evaluates them with black-box metrics, and updates a parameterized sampling distribution. Across a diverse UniProt protein set and the SARS-CoV-2 spike protein, our method consistently improves the chosen objective, with particularly strong gains on average unpaired probability and accessible uridine percentage compared to LinearDesign and EnsembleDesign. Moreover, our multi-objective COMBO formulation enables weight-controlled exploration of the design space and naturally extends to incorporate additional computable metrics.}


\maketitle

\section{Introduction}
\label{sec:intro}
% !TEX root = main.tex

%\section{Introduction}
Designing optimized messenger RNA (mRNA) sequences has received increasing attention in recent years, driven in part by the success of mRNA vaccines during the COVID-19 pandemic~\cite{baden2021mrna1273,polack2020bnt162b2}. Given a target protein sequence, mRNA design seeks a synonymous coding sequence that preserves translation while optimizing desired mRNA properties. Because codon degeneracy creates an exponentially large set of synonymous sequences, exhaustive search is infeasible~\cite{zhang2023lineardesign}; effective methods must therefore navigate this space under explicitly defined objectives.

Several methods have demonstrated that large-scale optimization over synonymous spaces is feasible when objectives are clearly defined. \lineardesign optimized synonymous mRNAs under a minimum free energy (\MFE) objective using dynamic programming, enabling efficient search at scale~\cite{zhang2023lineardesign}. More recently, \ensembledesign~\cite{dai2025ensembledesign}
and
JAX-RNAfold~\cite{krueger+ward:2025}
optimized a distribution-level objective, ensemble free energy (\EFE),
using continuous optimization,
but with very different approaches
(e.g., the former extended LinearDesign's lattice parsing framework
while the latter did not).
These approaches highlight the value of objective-driven optimization for navigating the vast synonymous coding space. However, practical mRNA design involves many objectives beyond \MFE and \EFE, and different applications often require different trade-offs. One example is the average unpaired probability (\AUP), which summarizes expected unpairedness across the transcript and is often used as a proxy related to degradation~\cite{leppek2022combinatorial}. Another example is accessible uridine percentage (accessible U\% / \AccessU), which we introduce in this work as a user-defined objective that measures the fraction of uridines that are structurally accessible (i.e., more likely to be unpaired). These considerations motivate a more general optimization framework that can directly optimize diverse objectives, rather than focusing on a small set of classical targets.

Motivated by this gap, we propose a general-purpose framework for mRNA design based on sampling-based continuous optimization. Our sampling-based optimization perspective is inspired by SamplingDesign, which demonstrates that continuous optimization coupled with Monte-Carlo sampling can be an effective paradigm for RNA design~\cite{tang2024samplingdesign}. Our method maintains a continuously parameterized sampling distribution over amino-acid--preserving nucleotide sequences and improves the distribution iteratively using samples scored by external evaluators. Concretely, each iteration draws candidate sequences from the current distribution, evaluates them under a chosen objective (for example, \MFE, \EFE, \CAI, \AUP, \AccessU, or weighted combinations of multiple metrics such as \COMBO), and updates the distribution parameters using sampling-based gradient estimates computed from the sampled sequences and their objective scores. This formulation supports optimizing both established objectives and additional potential mRNA evaluation objectives.

We evaluate on a diverse set of proteins from UniProt spanning a wide length range and on the SARS-CoV-2 spike protein as a representative long target, using a consistent scoring pipeline. Under single-metric optimization, our method consistently improves the targeted objective; most notably, when optimizing \AUP or \AccessU, our designed sequences achieve lower values than both \lineardesign and \ensembledesign. Under \COMBO optimization, our method enables weight-controlled navigation of the design space to obtain sequences that satisfy different optimization preferences. Importantly, because the optimization loop treats metrics as black-box evaluators of sampled sequences, the framework naturally extends to incorporate additional computable metrics beyond the current set, enabling richer objective combinations for future mRNA design tasks.

\vspace{-0.5cm}
\section{Preliminaries: mRNA Design for Protein}
\label{sec:prelim}
% !TEX root = main.tex

%\section{Preliminaries}
\subsection{Protein Encoding in mRNA}
\label{sec:prelim-background}

Let $\vecp=(\ptoken{1},\ptoken{2},\ldots,\ptoken{\protlen})$ be a target protein sequence of length $\protlen$,
where each $\ptoken{i}$ is an amino acid.
Let $\N \defeq \Nset$ denote the RNA nucleotide alphabet.
An mRNA coding sequence is a nucleotide string
$\vecx=(\xtoken{1},\xtoken{2},\ldots,\xtoken{\rnalen})\in\N^{\rnalen}$,
where $\xtoken{i}\in\N$ and $\rnalen = 3\protlen$.

Partition $\vecx$ into $\protlen$ consecutive codons (triplets) in $\N^{3}$.
Under the standard genetic code, each codon maps to an amino acid (or a stop signal);
concatenating the resulting amino acids yields a protein sequence.
We say that $\vecx$ encodes $\vecp$ if this translation yields $\vecp$.
Since multiple codons can map to the same amino acid, there are typically many distinct
mRNA sequences that encode the same protein.
Accordingly, our design space consists of synonymous coding sequences for $\vecp$:
all candidates encode $\vecp$ but may differ at the nucleotide level.

\subsection{mRNA Design Problem}
\label{sec:problem}

Given a fixed target protein sequence $\vecp$, mRNA design seeks a coding sequence $\vecx$
that encodes $\vecp$ while optimizing sequence-level properties measured by computable metrics.

\paragraph{Design space.}
We define the synonymous design space for $\vecp$ as
\begin{equation}
\designspaceof{\vecp}
\;\defeq\;
\left\{\vecx\in\N^{\rnalen}\;:\;\translate(\vecx)=\vecp\right\},
\end{equation}
i.e., the set of all mRNA coding sequences that translate to $\vecp$ under the standard genetic code.

\paragraph{Objective.}
Although all $\vecx\in\designspaceof{\vecp}$ encode the same protein, different synonymous choices
can yield different properties relevant to downstream use.
We assess a candidate by different computable metrics
$\rawmetrick{1}(\vecx,\vecp),\ldots,\rawmetrick{\metricindex}(\vecx,\vecp)$, where $\metricindex\in\{1,\ldots,\numMetrics\}$.
Because some metrics are to be maximized and others minimized, we map each raw metric to a
direction-aligned term $\metrick{\metricindex}(\vecx,\vecp)$ via a monotone transformation
\begin{equation}
\metrick{\metricindex}(\vecx,\vecp)\;\defeq\;\metrictfk{k}\!\big(\rawmetrick{\metricindex}(\vecx,\vecp)\big),
\end{equation}
so that smaller $\metrick{\metricindex}$ always indicates a better score, i.e., better performance according to metric $k$.

We then optimize a scalar objective over $\designspaceof{\vecp}$ using a normalized weighted sum of objective terms:
\begin{equation}
\Obj(\vecx,\vecp)
\;\defeq\;
\sum_{\metricindex=1}^{\numMetrics} \weightk{\metricindex}\,\metrick{\metricindex}(\vecx,\vecp),
\qquad
\weightk{\metricindex}\ge 0,\ \ \sum_{\metricindex=1}^{\numMetrics} \weightk{\metricindex}=1.
\end{equation}
When $\numMetrics=1$, this reduces to optimizing a single metric.
The design problem is
\begin{equation}
\optx
\;\in\;
\argmin_{\vecx\in\designspaceof{\vecp}}
\Obj(\vecx,\vecp).
\end{equation}

\subsection{Sequence Optimization Metrics}
\label{sec:metrics}

For a fixed target protein $\vecp$, each candidate coding sequence
$\vecx\in\designspaceof{\vecp}$ encodes the same protein but can exhibit different
sequence-level and structure-level behaviors relevant to downstream use.
We therefore evaluate $\vecx$ with several standard computable metrics.

\noindent\textbf{Folding model and notation.}
RNA secondary-structure metrics are based on the standard thermodynamic view
that a sequence $\vecx$ can adopt many feasible secondary structures, each
associated with an energy.
Let $\StructSet{\vecx}$ be the set of feasible secondary structures for $\vecx$
under a chosen energy model, and let $\freeenergy(\vecx,\vecy)$ denote the Gibbs
free energy of structure $\vecy\in\StructSet{\vecx}$.
We use thermodynamic parameters $(\GasConst,\Temp)$, where $\GasConst$ is the universal gas constant
and $\Temp$ is temperature.
Let $\bpp{i}{j}(\vecx)$ denote the marginal probability that positions $(i,j)$ form
a base pair in the Boltzmann ensemble, and define the unpaired probability of position $i$
\begin{equation}
\unp{i}(\vecx)\;\defeq\;1-\sum_{j\neq i} \bpp{i}{j}(\vecx),
\qquad i=1,\ldots,\rnalen.
\label{eq:unpaired_prob}
\end{equation}

\paragraph{Minimum free energy (\MFE).}
The minimum free energy summarizes the most stable (lowest-energy) structure:
\begin{equation}
\MFE(\vecx)\;\defeq\;\min_{\vecy\in\StructSet{\vecx}} \freeenergy(\vecx,\vecy).
\label{eq:mfe_def}
\end{equation}
We compute \MFE using \linearfold~\cite{huang2019linearfold} in this work.

\paragraph{Boltzmann ensemble and ensemble free energy (\EFE).}
Rather than focusing on a single structure, the partition function aggregates
contributions from all feasible structures, weighting each by its Boltzmann
factor:
\begin{equation}
\PF(\vecx)\;\defeq\;\sum_{\vecy\in\StructSet{\vecx}}
\exp\!\left(-\frac{\freeenergy(\vecx,\vecy)}{\RT}\right).
\label{eq:pf_def}
\end{equation}
The corresponding ensemble free energy is
\begin{equation}
\EFE(\vecx)\;\defeq\;-\RT\,\log \PF(\vecx),
\label{eq:efe_def}
\end{equation}
which we compute using \linearpartition~\cite{zhang2020linearpartition} in this work.

\paragraph{Average unpaired probability (\AUP).}
The average unpaired probability is
\begin{equation}
\AUP(\vecx)\;\defeq\;\frac{1}{\rnalen}\sum_{i=1}^{\rnalen} \unp{i}(\vecx).
\label{eq:aup}
\end{equation}

\paragraph{Accessible U\% (\AccessU).}
Accessible U\% weights unpaired probabilities by whether the nucleotide is U:
\begin{equation}
\AccessU(\vecx)\;\defeq\;\frac{1}{\rnalen}\sum_{i=1}^{\rnalen}\one[\xtoken{i}=\nucU]\;\unp{i}(\vecx).
\label{eq:accu}
\end{equation}

\paragraph{Codon Adaptation Index (\CAI).}
\CAI measures synonymous codon-usage bias relative to a reference
set of highly expressed genes, and is commonly used as a metric of codon
optimality~\cite{sharp1987cai}.

\paragraph{Combined objective (\COMBO).}
To express trade-offs among multiple criteria, we combine direction-aligned
terms in a normalized weighted sum.
Since $\CAI(\vecx)$, $\AUP(\vecx)$, and $\AccessU(\vecx)$ lie in $[0,1]$, they are
directly comparable after aligning directions (maximized terms are converted to
minimization by $1-\cdot$).
In contrast, $\MFE(\vecx)$ is length-dependent; we therefore normalize it using
a fixed baseline sequence for the same target protein.
Let $\MFE_{\mathrm{LD}}(\vecp)$ denote the \MFE of the \lineardesign{} baseline
sequence for $\vecp$.
We use the normalized MFE score
\begin{equation}
\mfe(\vecx)\;\defeq\;\frac{\MFE(\vecx)}{\MFE_{\mathrm{LD}}(\vecp)},
\label{eq:norm_mfe}
\end{equation}
so that larger $\mfe(\vecx)$ indicates a better (more stable) design relative to
the baseline.

Putting these together, the combined objective is
\begin{equation}
\begin{aligned}
\COMBO(\vecx,\vecp)\;\defeq\;&
\alphaw\bigl(1-\CAI(\vecx)\bigr)
+\betaw\,\AUP(\vecx) \\
&+\gammaw\,\AccessU(\vecx)
+\deltaw\bigl(1-\mfe(\vecx)\bigr)
\end{aligned}
\label{eq:combo}
\end{equation}
where $\alphaw,\betaw,\gammaw,\deltaw\ge 0$ and $\alphaw+\betaw+\gammaw+\deltaw=1$.

\vspace{-0.4cm}
\section{Parameterized Sampling Lattice for Synonymous mRNA Design}
\label{sec:lattice}
%\section{Parameterized sampling lattice for synonymous mRNA design}
\begin{figure*}[t]
\centering
\begin{tikzpicture}

% -------- Top row: two PDFs (left edges fixed) --------
\node (P1) [anchor=north west] at (0,0)
  {\includegraphics[width=0.50\textwidth]{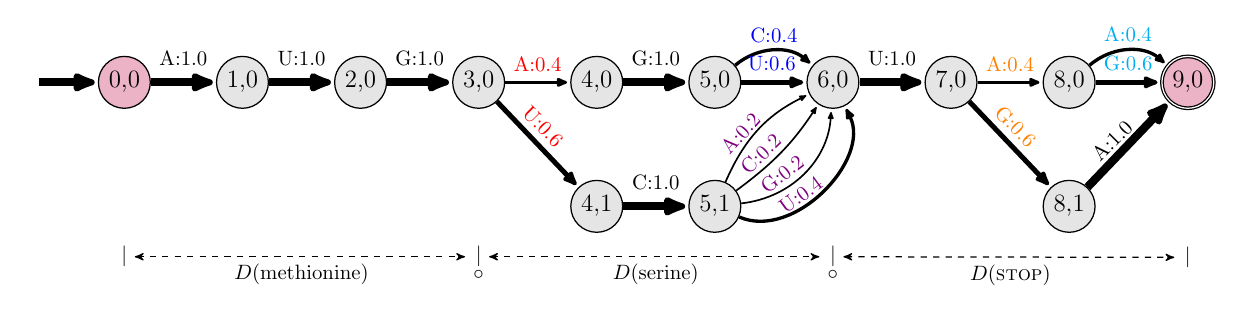}};

\node (P2) [anchor=north west] at ($(P1.north east)+(0,0)$)
  {\includegraphics[width=0.50\textwidth]{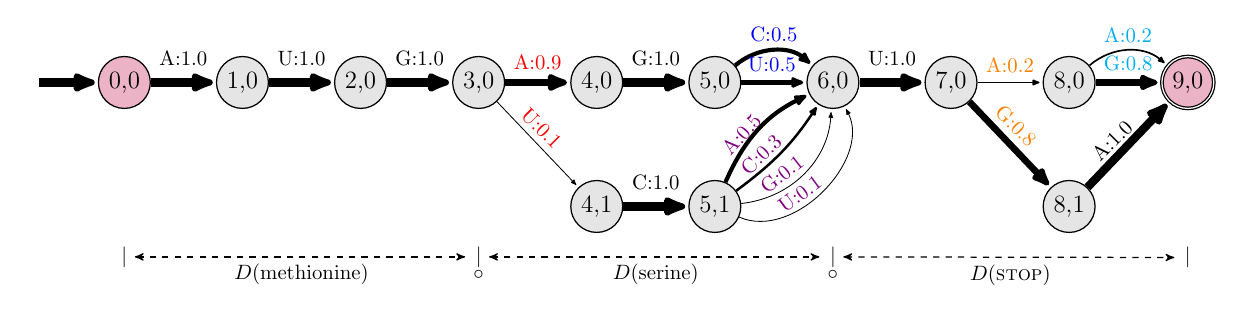}};

\node[anchor=north west] at ($(P1.north west)+(0,2mm)$) {\small (a) Initial distribution lattice};
\node[anchor=north west] at ($(P2.north west)+(0,2mm)$) {\small (b) Updated distribution lattice};

% -------- Bottom row: two LaTeX tables (left edges aligned to plots) --------
\node (T1) [anchor=north] at ($(P1.south)+(0,-8mm)$) {%
  \begin{minipage}[t]{0.46\textwidth}\vspace{0pt}
    \centering\small
    \resizebox{\linewidth}{!}{%
        \begin{tabular}{@{} >{\ttfamily}l
                        S[table-format=-1.6]
                        S[table-format=2.2]
                        S[table-format=2.2] @{}}
          \toprule
          {\normalfont\textbf{mRNA}} &
          {\normalfont\textbf{EFE} (kcal/mol)} &
          {\normalfont\textbf{AUP} (\%)} &
          {\normalfont\textbf{accessible U} (\%)} \\
          \midrule
            AUGUCUUAA & –0.000942 & 99.96 & 44.43 \\
            AUGAGCUGA & –0.000597 & 99.98 & 22.21 \\
            AUGUCUUAG & –0.000737 & 99.97 & 44.43 \\
            AUGAGUUGA & –0.000778 & 99.97 & 33.32 \\
            AUGUCUUAA & –0.000942 & 99.96 & 44.43 \\
            AUGUCUUGA & –0.000849 & 99.97 & 44.43 \\
            AUGUCCUGA & –0.000654 & 99.98 & 33.32 \\
            AUGAGUUAG & –0.000568 & 99.98 & 33.32 \\
            AUGUCGUGA & –0.000848 & 99.96 & 33.32 \\
            AUGUCGUGA & –0.000848 & 99.96 & 33.32 \\
          \bottomrule
        \end{tabular}
    }
    
    \captionof{table}{Initial sample of 10 mRNAs based on the initial distribution.}
  \end{minipage}
};

\node (T2) [anchor=north] at ($(P2.south)+(0,-8mm)$) {%
  \begin{minipage}[t]{0.46\textwidth}\vspace{0pt}
    \centering\small
    \resizebox{\linewidth}{!}{%
        \begin{tabular}{@{} >{\ttfamily}l
                        S[table-format=-1.6]
                        S[table-format=2.2]
                        S[table-format=2.2] @{}}
          \toprule
          {\normalfont\textbf{mRNA}} &
          {\normalfont\textbf{EFE} (kcal/mol)} &
          {\normalfont\textbf{AUP} (\%)} &
          {\normalfont\textbf{accessible U} (\%)} \\
          \midrule
            AUGAGCUGA & -0.000597 & 99.98 & 22.21 \\
            AUGAGUUGA & -0.000778 & 99.97 & 33.32 \\
            AUGAGCUGA & -0.000597 & 99.98 & 22.21 \\
            AUGAGUUGA & -0.000778 & 99.97 & 33.32 \\
            AUGAGCUAG & -0.000387 & 99.98 & 22.21 \\
            AUGAGCUGA & -0.000597 & 99.98 & 22.21 \\
            AUGAGUUGA & -0.000778 & 99.97 & 33.32 \\
            AUGAGUUGA & -0.000778 & 99.97 & 33.32 \\
            AUGAGCUGA & -0.000597 & 99.98 & 22.21 \\
            AUGAGUUGA & -0.000778 & 99.97 & 33.32 \\
          \bottomrule
        \end{tabular}
    }

    \captionof{table}{New sample of 10 mRNAs based on the updated distribution.}
  \end{minipage}
};

% -------- Arrows --------
\draw[->,thick] ($(P1.south)+(0,-1mm)$) -- node[right]{\small Sample and Evaluate} ($(T1.north)+(0,1mm)$);
\draw[->,thick] ($(P2.south)+(0,-1mm)$) -- node[right]{\small Sample and Evaluate} ($(T2.north)+(0,1mm)$);
% \draw[->,thick] ($(P1.east)+(1mm,0)$) -- node[above]{\small update} ($(P2.west)+(-1mm,0)$);

% Left-bottom (T1) -> Right-top (P2): diagonal arrow + sloped label
\draw[->,thick]
  ($(T1.east)+(0,0)$) -- node[above,sloped]{\small Gradient Update}
  ($(P2.west)+(5mm,-6mm)$);

% LEFT

\draw[rounded corners=2pt, draw=cyan, line width=0pt, fill=cyan, fill opacity=0.10]
  ($(P1.south west)!0.66!(P1.north west) + (36.0mm,0mm)$) circle (3.2mm);

\draw[rounded corners=2pt, draw=cyan, line width=0pt, fill=cyan, fill opacity=0.10]
  ($(P1.south west)!0.30!(P1.north west) + (48mm,0mm)$) rectangle
  ($(P1.south west)!0.90!(P1.north west) + (54mm,0mm)$);

\draw[rounded corners=2pt, draw=cyan, line width=0pt, fill=cyan, fill opacity=0.10]
  ($(T1.south west)!0.11!(P1.north west) + (52mm,0mm)$) rectangle
  ($(T1.south west)!0.60!(P1.north west) + (75mm,0mm)$);

% RIGHT

\draw[rounded corners=2pt, draw=cyan, line width=0pt, fill=cyan, fill opacity=0.10]
  ($(P2.south west)!0.66!(P2.north west) + (36.0mm,0mm)$) circle (3.2mm);

\draw[rounded corners=2pt, draw=cyan, line width=0pt, fill=cyan, fill opacity=0.10]
  ($(P2.south west)!0.30!(P2.north west) + (48mm,0mm)$) rectangle
  ($(P2.south west)!0.90!(P2.north west) + (54mm,0mm)$);

\draw[rounded corners=2pt, draw=cyan, line width=0pt, fill=cyan, fill opacity=0.10]
  ($(T2.south west)!0.11!(P2.north west) + (52mm,0mm)$) rectangle
  ($(T2.south west)!0.60!(P2.north west) + (75mm,0mm)$);

\end{tikzpicture}

% \caption{Your overall caption here.}
\caption{\textbf{The parameterized sampling lattice and an illustrative update--sample--evaluate iteration.}
\textit{(a--b) parameterized sampling lattice.} To avoid enumerating the exponentially large synonymous space for a fixed protein $\vecp$, we represent all valid coding sequences as a DFA-based lattice, where each complete path corresponds to a synonymous mRNA $\vecx$. We then equip the lattice with probabilistic parameters, forming a pDFA in which each state defines a locally normalized distribution over its outgoing edges (Eq.~\ref{eq:local_norm}). Edge labels are nucleotides, and sampling generates an mRNA by traversing the lattice and concatenating the emitted labels.
\textbf{Illustrative workflow.} Starting from an initialized lattice (a), we sample candidate mRNAs and Evaluate them under the chosen objective (here, minimizing accessible U\%). Using these scores, we perform a gradient update on the lattice parameters, yielding an updated probabilistic lattice (b), from which we sample and evaluate again. After the update, the sampled sequences exhibit a clear reduction in accessible U\% (as highlighted by the cyan regions in the table), and the corresponding decision region shifts probability mass away from U-emitting branches (as highlighted by the cyan regions in the pDFA), consistent with reducing U-rich choices as one effective route to lowering accessible U\%. This update--sample--evaluate loop is repeated until the optimization metrics converge.}

\label{fig:lattice_overview}
\end{figure*}

The synonymous design space $\designspaceof{\vecp}$ is exponentially large and
is defined by a hard amino-acid constraint.
To avoid enumerating $\designspaceof{\vecp}$ while preserving feasibility, we
represent it as a lattice and parameterize a sampling distribution directly on
this structure (Fig.~\ref{fig:lattice_overview}).

\paragraph{DFA lattice representation.}
Following \lineardesign, the set of coding sequences that encode a fixed target
protein $\vecp$ is compactly represented by a deterministic finite-state
automaton (DFA)~\citep{zhang2023lineardesign}.
In this lattice, each edge is labeled by a nucleotide, and any complete path
generates an mRNA sequence $\vecx\in\N^{\rnalen}$ by concatenating edge labels.
By construction, every complete path translates to $\vecp$, and conversely,
every $\vecx\in\designspaceof{\vecp}$ is represented by at least one complete
path in the lattice.
This DFA therefore provides a constraint-preserving representation of the
synonymous space.

\paragraph{From a lattice to a sampling distribution.}
To enable continuous optimization in an otherwise discrete space, we follow the
continuous relaxation perspective of \ensembledesign and equip the same lattice
with probabilistic parameters, yielding a probabilistic DFA (pDFA) that defines
a distribution over synonymous sequences~\citep{dai2025ensembledesign}.
Crucially, probability mass is assigned only to valid DFA paths, so sampling
from the pDFA produces sequences in $\designspaceof{\vecp}$ automatically.

Formally, let $\latticestate$ denote a lattice state and let $\outedges{\latticestate}$ be its set of
outgoing edges.
We associate each state with a categorical distribution over outgoing edges,
\begin{equation}
\transprob{\vectheta}{\edgeact}{\latticestate}\ge 0,
\qquad
\sum_{\edgeact\in\outedges{\latticestate}} \transprob{\vectheta}{\edgeact}{\latticestate}=1,
\label{eq:local_norm}
\end{equation}
where $\vectheta$ collects all trainable parameters.
Sampling corresponds to traversing the lattice by repeatedly selecting an
outgoing edge $\edgeact\sim \transprob{\vectheta}{\cdot}{\latticestate}$ and advancing to the next state.
The resulting concatenation of edge labels yields a sampled
$\vecx\sim \seqdist{\vectheta}(\cdot)$ supported on $\designspaceof{\vecp}$.

\paragraph{Parameterized local transitions.}
Rather than optimizing probabilities directly on the simplex, we introduce
unconstrained logits $\logit{\latticestate}{\edgeact}\in\mathbb{R}$ for each $\edgeact\in\outedges{\latticestate}$.
Local transition probabilities are obtained via the softmax,
\begin{equation}
\transprob{\vectheta}{\edgeact}{\latticestate}
=
\frac{\exp(\logit{\latticestate}{\edgeact})}{\sum_{\edgeact'\in\outedges{\latticestate}} \exp(\logit{\latticestate}{\edgeact'})},
\qquad \edgeact\in\outedges{\latticestate},
\label{eq:softmax}
\end{equation}
which automatically enforces nonnegativity and normalization while allowing
updates to be performed in an unconstrained Euclidean space.

\paragraph{Sequence probability and path factorization.}
A sampled sequence $\vecx$ corresponds to a complete path
\(
\pathseq=(\latticestate_0,\edgeact_0,\ldots,\latticestate_{\rnalen-1},\edgeact_{\rnalen-1},\latticestate_{\rnalen})
\)
on the lattice, where $\edgeact_\posindex$ denotes the chosen outgoing edge at state $\latticestate_\posindex$.
Under~\eqref{eq:softmax}, the probability of sampling the path (and thus the
sequence) factorizes along the path:
\begin{equation}
\seqdist{\vectheta}(\vecx)=\seqdist{\vectheta}(\pathseq)=\prod_{\posindex=0}^{\rnalen-1} \transprob{\vectheta}{\edgeact_\posindex}{\latticestate_\posindex}.
\label{eq:path_prob}
\end{equation}
Equivalently,
\begin{equation}
\log \seqdist{\vectheta}(\vecx)=\sum_{\posindex=0}^{\rnalen-1}\log \transprob{\vectheta}{\edgeact_\posindex}{\latticestate_\posindex}.
\label{eq:path_logprob}
\end{equation}

This factorization is central to our approach.
Our objective $\COMBO(\vecx,\vecp)$ is evaluated on discrete sequences using
folding-based metrics and is treated as a black-box function of $\vecx$.
Although $\COMBO$ is not differentiable with respect to nucleotide identities,
the pDFA provides a differentiable handle through $\log \seqdist{\vectheta}(\vecx)$.
In the next section, we leverage~\eqref{eq:path_logprob} to derive
sampling-based gradient estimates and implement a sample--evaluate--update
procedure to optimize $\vectheta$.

\vspace{-0.5cm}
\section{Sampling-Based Continuous Optimization on the Lattice}
\label{sec:algorithm}
%\section{Sampling-based continuous optimization on the lattice}

In this section, we present our sampling-based continuous optimization
algorithm for synonymous mRNA design.
The method iterates a sample--evaluate--update loop on the parameterized
sampling lattice, using a score-function gradient estimator in logit space and
terminating via an early-stopping rule.

\subsection{Sampling}
\label{subsec:sampling}

Starting from the initial state, a candidate sequence $\vecx$ is generated by
traversing the lattice and repeatedly sampling an outgoing edge
$\edgeact_{\posindex}\sim \transprob{\vectheta}{\cdot}{\latticestate_{\posindex}}$.
Concatenating the nucleotide labels along the sampled path yields
$\vecx\in\N^{\rnalen}$.
For each sampled $\vecx$, we record the corresponding lattice path
\(
\pathseq
=
(\latticestate_0,\edgeact_0,\ldots,\latticestate_{\rnalen-1},\edgeact_{\rnalen-1},\latticestate_{\rnalen})
\),
where $\edgeact_{\posindex}$ is the chosen outgoing edge at
$\latticestate_{\posindex}$.

\subsection{Evaluation}
\label{subsec:evaluation}

Each sampled sequence $\vecx$ is assigned a scalar objective value
$\Obj(\vecx,\vecp)$, where $\Obj(\vecx,\vecp)$ is either a single metric or the
weighted combination $\COMBO(\vecx,\vecp)$ defined in
Section~\ref{sec:metrics}.
These objective values provide the learning signal for updating $\vectheta$.

\subsection{Gradient-Based Update}
\label{subsec:update}

The update step adjusts the lattice logits (and hence the induced sampling
distribution) to reduce the expected objective under the current lattice
distribution. 
We index optimization iterations by $\iterindex\in\{1,2,\ldots\}$, and let
$\vectheta_{\iterindex}$ denote the lattice parameters used to sample the batch
at iteration $\iterindex$ (i.e., after $\iterindex-1$ updates).
At iteration $\iterindex$, we perform a gradient-based update
\begin{equation}
\vectheta_{\iterindex+1}
=
\vectheta_{\iterindex}
-\stepsize\,\gradest_{\iterindex},
\label{eq:gd_update}
\end{equation}
where $\gradest_{\iterindex}$ is a stochastic estimate of
$\nabla_{\vectheta}\ExpObj(\vectheta_{\iterindex};\vecp)$ derived below and
$\stepsize$ is the learning rate. 
In practice, we use Adam to compute an adaptive update from $\gradest_{\iterindex}$
(Paragraph~\textbf{(vii)}).

% =========================
\paragraph{(i) Expected objective under the lattice distribution.}
Fix a target protein $\vecp$.
Sampling on the lattice induces a distribution
$\seqdist{\vectheta}(\vecx)$ over synonymous coding sequences $\vecx$.
We minimize the expected objective
\begin{equation}
\ExpObj(\vectheta;\vecp)
\;\defeq\;
\E_{\vecx\sim \seqdist{\vectheta}(\cdot)}\!\left[\Obj(\vecx,\vecp)\right].
\label{eq:Jtheta_25}
\end{equation}

% =========================
\paragraph{(ii) Score-function (log-derivative) gradient.}
Since $\vecx$ is discrete, we differentiate~\eqref{eq:Jtheta_25} using the
log-derivative trick:
\begin{align}
\nabla_{\vectheta}\ExpObj(\vectheta;\vecp)
&=
\nabla_{\vectheta} \sum_{\vecx} \seqdist{\vectheta}(\vecx)\,\Obj(\vecx,\vecp) \notag\\
&=
\sum_{\vecx} \seqdist{\vectheta}(\vecx)\,\Obj(\vecx,\vecp)\,
\nabla_{\vectheta}\log \seqdist{\vectheta}(\vecx) \notag\\
&=
\E_{\vecx\sim \seqdist{\vectheta}(\cdot)}\!\left[
\Obj(\vecx,\vecp)\,\nabla_{\vectheta}\log \seqdist{\vectheta}(\vecx)
\right].
\label{eq:score_grad_25}
\end{align}
With $\batchsize$ i.i.d.\ samples
$\vecx^{(1)},\ldots,\vecx^{(\batchsize)}\sim \seqdist{\vectheta}(\cdot)$,
a Monte Carlo estimate of~\eqref{eq:score_grad_25} is
\begin{equation}
\widehat{\nabla_{\vectheta}\ExpObj}
=
\frac{1}{\batchsize}\sum_{\sampleindex=1}^{\batchsize}
\Obj(\vecx^{(\sampleindex)},\vecp)\,
\nabla_{\vectheta}\log \seqdist{\vectheta}\!\big(\vecx^{(\sampleindex)}\big).
\label{eq:mc_global_25}
\end{equation}
The remaining task is to express
$\nabla_{\vectheta}\log \seqdist{\vectheta}\!\big(\vecx^{(\sampleindex)}\big)$
in terms of local lattice decisions.

% =========================
\paragraph{(iii) Path decomposition of the score term.}
Each sampled sequence $\vecx$ corresponds to a complete lattice path
\(
\pathseq
=
(\latticestate_0,\edgeact_0,\ldots,\latticestate_{\rnalen-1},\edgeact_{\rnalen-1},\latticestate_{\rnalen})
\)
of length $\rnalen$.
By the path factorization in Section~\ref{sec:lattice}
(Eqs.~\eqref{eq:path_prob}--\eqref{eq:path_logprob}),
\begin{equation}
\nabla_{\vectheta} \log \seqdist{\vectheta}(\vecx)
=
\sum_{\posindex=0}^{\rnalen-1}
\nabla_{\vectheta}\log \transprob{\vectheta}{\edgeact_{\posindex}}{\latticestate_{\posindex}}.
\label{eq:score_sum}
\end{equation}
Substituting~\eqref{eq:score_sum} into~\eqref{eq:mc_global_25} reduces the global
score term to a sum of local contributions along the sampled path.

% =========================
\paragraph{(iv) Local softmax derivative.}
At each visited state $\latticestate$, the outgoing distribution over
$\outedges{\latticestate}$ is parameterized by logits via the softmax in
Eq.~\eqref{eq:softmax}.
Let $\edgeact\in\outedges{\latticestate}$ be the chosen edge on the sampled path,
and let $\edgealt\in\outedges{\latticestate}$ index an arbitrary outgoing edge.
Then
\begin{equation}
\frac{\partial}{\partial \logit{\latticestate}{\edgealt}}
\log \transprob{\vectheta}{\edgeact}{\latticestate}
=
\ind{\edgealt=\edgeact}
-\transprob{\vectheta}{\edgealt}{\latticestate}.
\label{eq:local_logit_grad_25}
\end{equation}
This provides an explicit per-logit local score term.

% =========================
\paragraph{(v) Mean--variance normalization.}
For the batch scores
$\{\Obj(\vecx^{(\sampleindex)},\vecp)\}_{\sampleindex=1}^{\batchsize}$,
we compute the batch mean and standard deviation
\begin{equation}
\begin{aligned}
\baseline
&\defeq
\frac{1}{\batchsize}\sum_{\sampleindex=1}^{\batchsize}
\Obj(\vecx^{(\sampleindex)},\vecp),\\
\batchstd
&\defeq
\sqrt{\frac{1}{\batchsize}\sum_{\sampleindex=1}^{\batchsize}
\bigl(\Obj(\vecx^{(\sampleindex)},\vecp)-\baseline\bigr)^2}
+\vareps.
\end{aligned}
\label{eq:batch_stats_25}
\end{equation}

and define the normalized weight
\begin{equation}
\advw^{(\sampleindex)}
\;\defeq\;
\frac{\Obj(\vecx^{(\sampleindex)},\vecp)-\baseline}{\batchstd}.
\label{eq:advantage_25}
\end{equation}
Subtracting a baseline is a standard variance-reduction device for score-function
gradient estimators~\citep{williams1992reinforce,greensmith2004variance}.
Here we set $\baseline$ to the batch mean so that each sample is weighted by
its performance relative to the current batch rather than by the absolute scale
of $\Obj$.

% =========================
\paragraph{(vi) Fully expanded per-logit gradient estimator.}
Let the $\sampleindex$-th sampled sequence have path
\(
\pathseq^{(\sampleindex)}
=
(\latticestate^{(\sampleindex)}_0,\edgeact^{(\sampleindex)}_0,\ldots,
\latticestate^{(\sampleindex)}_{\rnalen-1},\edgeact^{(\sampleindex)}_{\rnalen-1},
\latticestate^{(\sampleindex)}_{\rnalen})
\).
Combining~\eqref{eq:mc_global_25}, \eqref{eq:score_sum}, and
\eqref{eq:local_logit_grad_25}, for any state $\latticestate$ and any
$\edgealt\in\outedges{\latticestate}$,
\begin{equation}
\begin{aligned}
\widehat{\frac{\partial \ExpObj(\vectheta;\vecp)}{\partial \logit{\latticestate}{\edgealt}}}
&=
\frac{1}{\batchsize}\sum_{\sampleindex=1}^{\batchsize} \advw^{(\sampleindex)}
\sum_{\posindex=0}^{\rnalen-1}
\ind{\latticestate^{(\sampleindex)}_{\posindex}=\latticestate}\,
\\
&\qquad\cdot
\Bigl(
\ind{\edgealt=\edgeact^{(\sampleindex)}_{\posindex}}
-\transprob{\vectheta}{\edgealt}{\latticestate}
\Bigr).
\end{aligned}
\label{eq:final_grad_expanded_25}
\end{equation}
The indicator $\ind{\latticestate^{(\sampleindex)}_{\posindex}=\latticestate}$
ensures that only states visited by the sampled paths contribute to the
corresponding logits.

% =========================
\paragraph{(vii) Adam updates in logit space.}
Let $\gradest_{\iterindex}$ denote the collection of all per-logit gradient
estimates in~\eqref{eq:final_grad_expanded_25} at iteration $\iterindex$.
We use the Adam optimizer, which maintains exponential moving averages of past
gradients and squared gradients to form a momentum term and an adaptive,
per-parameter step size~\citep{kingma2017adammethodstochasticoptimization}.
This is convenient for stochastic sampling-based gradient estimates and
typically yields more stable updates than a fixed learning rate.

We update logits with Adam:
\begin{align}
\adamm{\iterindex} &=
\beta_1\,\adamm{\iterindex-1} + (1-\beta_1)\,\gradest_{\iterindex}, \notag\\
\adamv{\iterindex} &=
\beta_2\,\adamv{\iterindex-1} + (1-\beta_2)\,\gradest_{\iterindex}^2, \notag\\
\adammhat{\iterindex} &=
\frac{\adamm{\iterindex}}{1-\beta_1^{\iterindex}},\qquad
\adamvhat{\iterindex} =
\frac{\adamv{\iterindex}}{1-\beta_2^{\iterindex}}, \notag\\
\vectheta_{\iterindex+1}
&=
\vectheta_{\iterindex}
-\stepsize \frac{\adammhat{\iterindex}}{\sqrt{\adamvhat{\iterindex}}+\adameps}.
\label{eq:adam_25}
\end{align}

\subsection{Early Stopping and Termination Criteria}
\label{sec:early_stop}

The sample--evaluate--update loop is stochastic and can be computationally
expensive, as each iteration requires evaluating folding-based objectives on a
batch of sequences.
In practice, the optimization often converges to a stable region of the search
space, after which additional iterations produce little or no improvement in
the observed objective values while incurring substantial runtime.
We therefore adopt an early stopping criterion that terminates the procedure
once progress on the target objective has plateaued.

\paragraph{Best-so-far tracking.}
At iteration $\iterindex$, we draw a batch
$\{\vecx^{(\sampleindex)}\}_{\sampleindex=1}^{\batchsize}\sim \seqdist{\vectheta_{\iterindex}}(\cdot)$
and compute their objective values
$\{\Obj(\vecx^{(\sampleindex)},\vecp)\}_{\sampleindex=1}^{\batchsize}$.
Let the current best objective value in the batch be
\begin{equation}
\batchbest_{\iterindex}
\;\defeq\;
\min_{\sampleindex\in\{1,\ldots,\batchsize\}}
\Obj(\vecx^{(\sampleindex)},\vecp),
\label{eq:batch_best}
\end{equation}
and let the best value observed up to iteration $\iterindex$ be updated by
\begin{equation}
\bestsofar_{\iterindex}
\;\defeq\;
\min\!\bigl(\bestsofar_{\iterindex-1},\,\batchbest_{\iterindex}\bigr),
\qquad
\bestsofar_{0}\;\defeq\;+\infty.
\label{eq:best_so_far}
\end{equation}

\paragraph{Patience-based early stopping.}
We declare an iteration to be improving if it achieves a sufficient
decrease relative to the previous best:
\begin{equation}
\bestsofar_{\iterindex}
\le
\bestsofar_{\iterindex-1}-\stopminimprove.
\label{eq:improvement_rule}
\end{equation}
Let $\patience$ be a patience window (in iterations). We stop if no improving
iteration occurs for $\patience$ consecutive updates, i.e., if the number of
iterations since the last improvement exceeds $\patience$.

\paragraph{Maximum-iteration safeguard.}
To bound runtime, we also impose a hard cap of $\maxiters$ iterations and stop
once $\iterindex=\maxiters$.

\medskip
For convenience, Algorithm~\ref{alg:opt_lattice} provides a compact summary of
the complete sample--evaluate--update procedure together with early stopping.
Fig.~\ref{fig:lattice_overview} complements this summary with an illustrative
example of how a single update shifts probability mass on the lattice and
changes the resulting batch of samples.

\begin{algorithm}[t]
\caption{Sampling-based continuous optimization}
\label{alg:opt_lattice}
\begin{algorithmic}[1]
\State \textbf{Input:} target protein $\vecp$, objective $\Obj(\cdot,\vecp)$, initialized lattice logits $\vectheta_{1}$
\State \textbf{Output:} best sequence $\vecx^\star$
\State Initialize optimizer state; $\vecx^\star\gets\emptyset$; $\bestsofar\gets+\infty$
\For{$\iterindex = 1,2,\ldots, \maxiters$}
  \State Sample a batch $\{\vecx^{(\sampleindex)},\pathseq^{(\sampleindex)}\}_{\sampleindex=1}^{\batchsize}\sim \seqdist{\vectheta_{\iterindex}}(\cdot)$
  \State Evaluate scores $\{\Obj(\vecx^{(\sampleindex)},\vecp)\}_{\sampleindex=1}^{\batchsize}$ and update $\vecx^\star,\bestsofar$
  \If{early-stopping criterion is met}
    \State \textbf{break}
  \EndIf
  \State Compute gradient estimate $\gradest_{\iterindex}$ and update logits to obtain $\vectheta_{\iterindex+1}$
\EndFor
\State \Return $\vecx^\star$
\end{algorithmic}
\end{algorithm}

\vspace{-0.4cm}
\section{Experimental Setup}
\label{sec:exp_setup}
%\section{Experimental Setup}

% =========================
% Figure: Single-metric evaluation composite
% =========================

\begin{figure*}[t]
\centering
\begin{tikzpicture}

% ---------- Row 1: three side-by-side plots ----------
\node (A) [anchor=north west] at (0,0)
  {\includegraphics[width=0.333\textwidth]{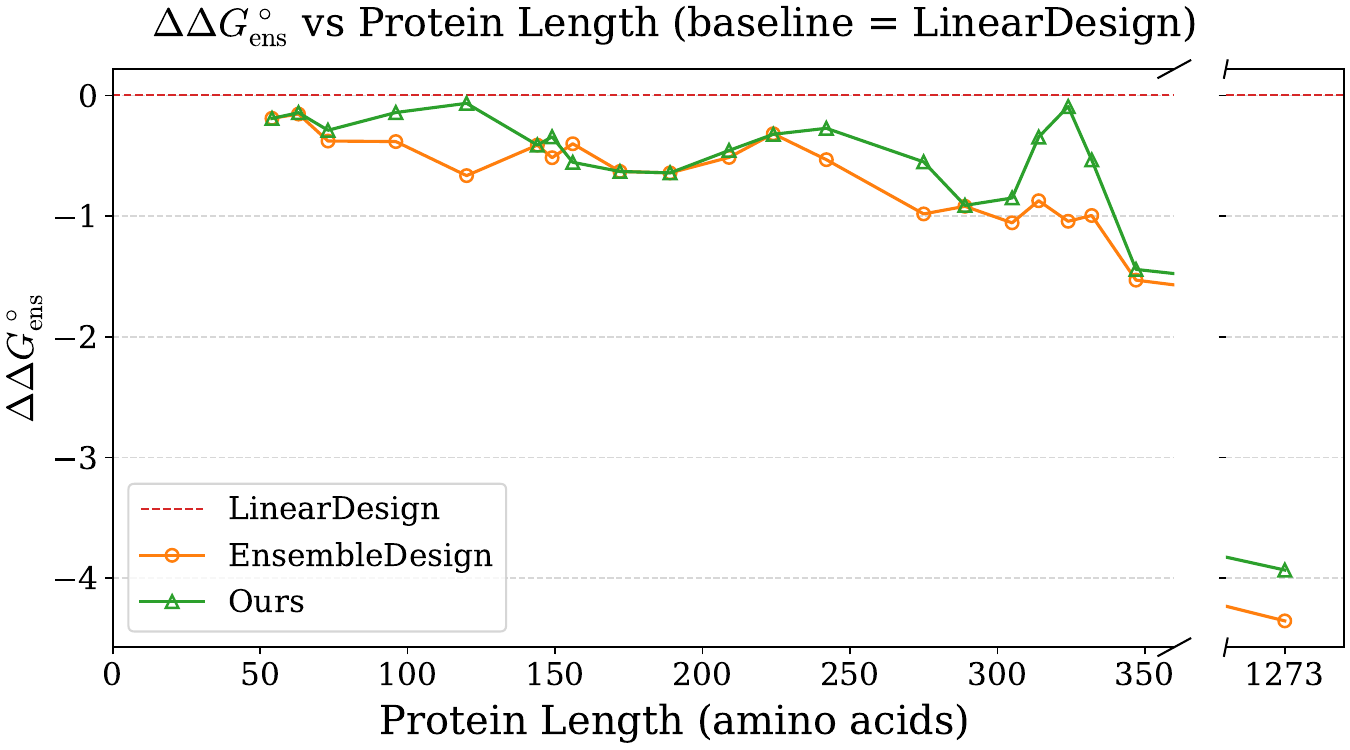}};
\node (B) [anchor=north west] at ($(A.north east)+(0,0)$)
  {\includegraphics[width=0.333\textwidth]{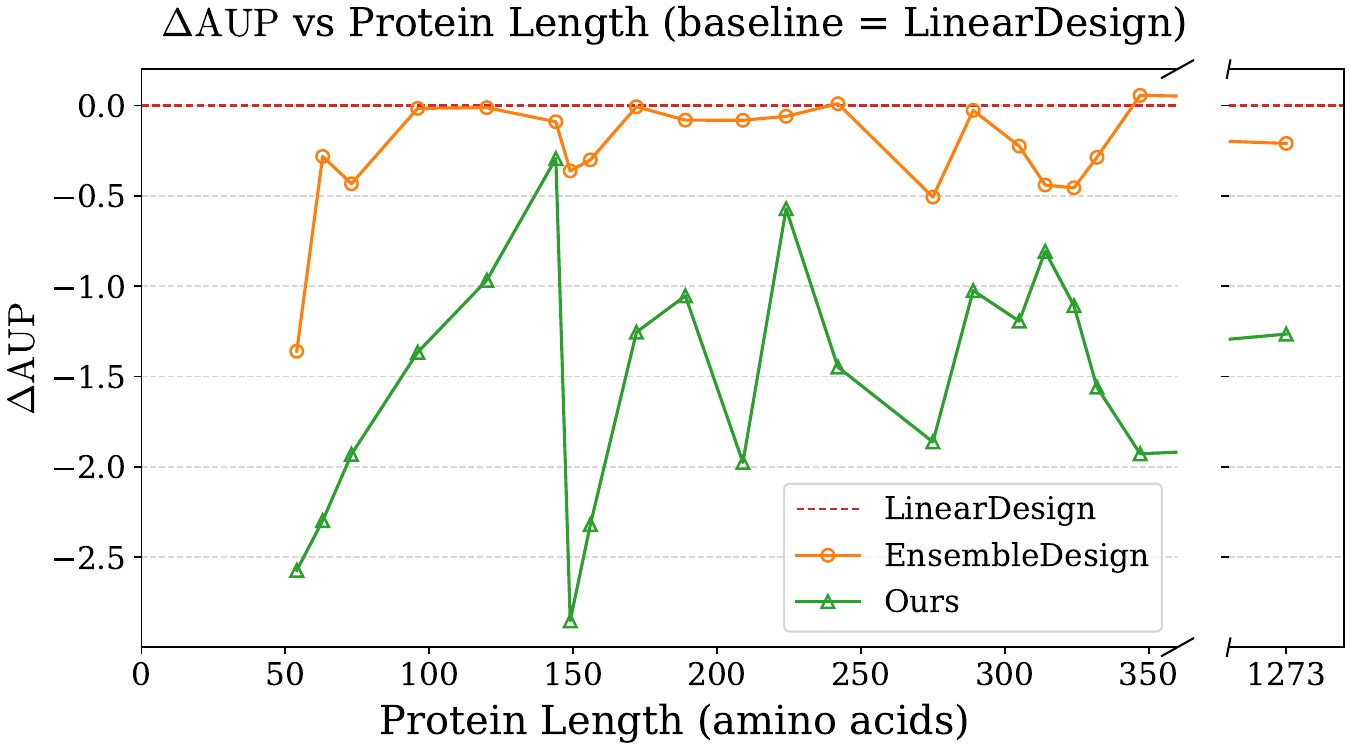}};
\node (C) [anchor=north west] at ($(B.north east)+(0,0)$)
  {\includegraphics[width=0.333\textwidth]{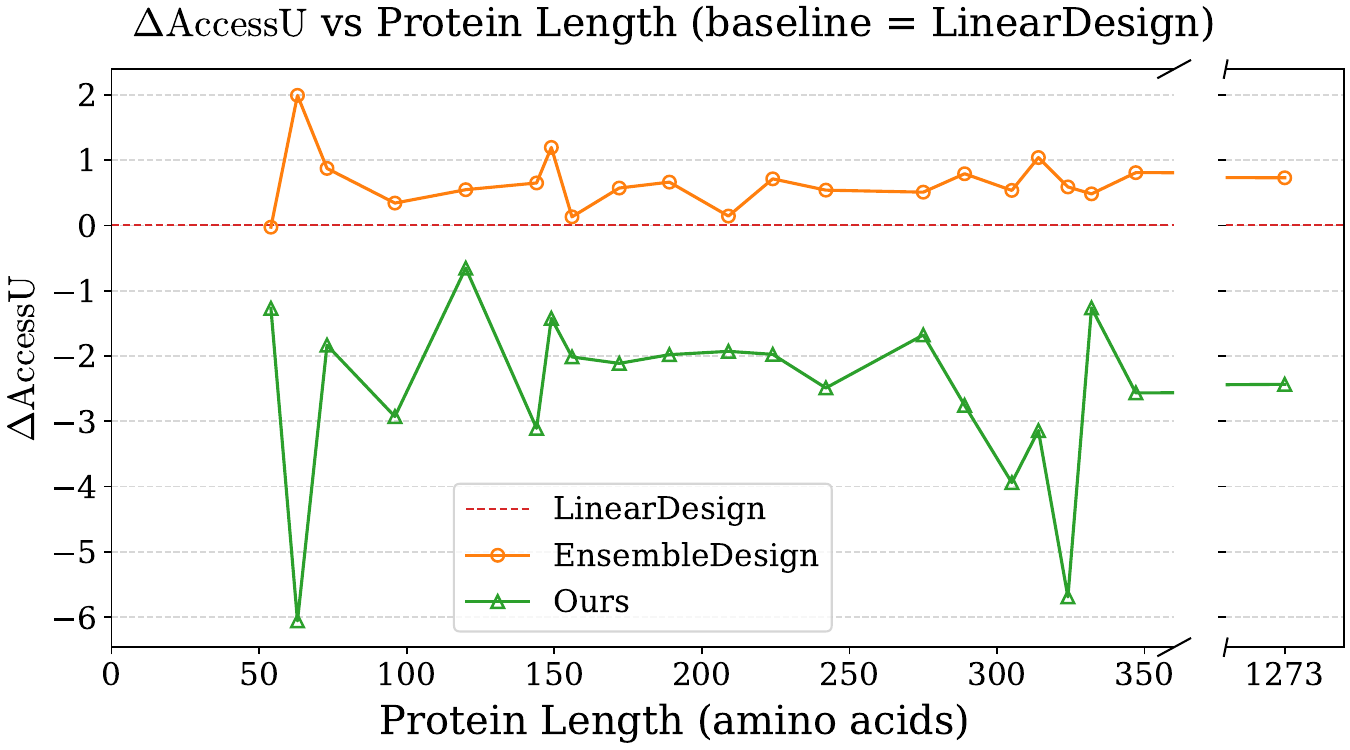}};

\node[anchor=south, align=center, text width=0.333\textwidth] at ($(A.north)+(0,1mm)$) {%
  \small (a) $\Delta\Delta G^{\circ}_{\mathrm{ens}}$ vs.\ protein length
};

\node[anchor=south, align=center, text width=0.333\textwidth] at ($(B.north)+(0,1mm)$) {%
  \small (b) $\Delta\mathrm{AUP}$ vs.\ protein length
};

\node[anchor=south, align=center, text width=0.333\textwidth] at ($(C.north)+(0,1mm)$) {%
  \small (c) $\Delta\mathrm{AccessU}$ vs.\ protein length
};

% ---------- Row 2: one wide table ----------
\node (T) [anchor=north] at ($(A.south)!0.5!(C.south)+(0,0mm)$) {%
  \begin{minipage}[t]{1.0\textwidth}\vspace{0pt}
    \centering\normalsize
    \setlength{\tabcolsep}{5pt}
    \renewcommand{\arraystretch}{1.10}
    
    \captionof{table}{\textbf{Single-metric optimization results across protein sequences.}
    This table compares \lineardesign, \ensembledesign, and our method on the UniProt set and the SARS-CoV-2 spike protein. For each protein, we report three metrics of the designed mRNA sequence: \EFE (denoted $\Delta G^{\circ}_{\mathrm{ens}}$), \AUP, and \AccessU, all computed under the shared scoring pipeline in Section~\ref{sec:exp_setup}. The \lineardesign block reports absolute metric values for its designed sequence. The \ensembledesign block reports metric differences relative to \lineardesign, $\Delta(\cdot)\defeq(\cdot)_{\mathrm{ED}}-(\cdot)_{\mathrm{LD}}$. The \textbf{Ours} block is also reported as differences relative to \lineardesign, but it aggregates three single-metric runs with distinct objectives (\EFE, \AUP, and \AccessU), and thus yields one optimized sequence per objective. The last three columns report the corresponding relative changes for the objective optimized in each run, indicated by $\objmark$.}
    
    \resizebox{0.98\linewidth}{!}{%
      \begin{tabular}{ | c c | c c c | c r c | c c c |}
      
      \toprule

      % ============================
      % UniProt block
      % ============================
      \multicolumn{11}{c}{\textbf{UniProt Proteins}} \\
      \cmidrule(lr){1-11}
      \addlinespace[0.6mm]

      \multicolumn{2}{c}{} &
      \multicolumn{3}{c}{\textbf{\lineardesign (LD)}} &
      \multicolumn{3}{c}{\textbf{\ensembledesign (vs. LD)}} &
      \multicolumn{3}{c}{\textbf{Ours (vs. LD)}} \\
      \cmidrule(lr){1-2}
      \cmidrule(lr){3-5}
      \cmidrule(lr){6-8}
      \cmidrule(lr){9-11}

      \textbf{ID} &
      \textbf{length (aa / nt)} &
      $\Delta G^{\circ}_{\mathrm{ens}}$ &
      $\AUP$ &
      $\AccessU$ &
      $\Delta\Delta G^{\circ}_{\mathrm{ens}}$ &
      $\Delta\AUP$ &
      $\Delta\AccessU$ &
      $\Delta\Delta G^{\circ}_{\mathrm{ens}}\objmark$ &
      $\Delta\AUP\objmark$ &
      $\Delta\AccessU\objmark$ \\
      \midrule

      % -------- UniProt rows --------
      \texttt{\textbf{Q13794}} & 54 / 162 & -113.39 & 20.64 & 2.33 & \textbf{-0.19} & -1.36 & -0.03 & \textbf{-0.19} & \textbf{-2.57} & \textbf{-1.27} \\
      \texttt{\textbf{Q9UI25}} & 63 / 189 & -126.07 & 25.87 & 6.46 & \textbf{-0.15} & -0.28 & 1.99 & -0.14 & \textbf{-2.30} & \textbf{-6.06} \\
      \texttt{\textbf{Q9BZL1}} & 73 / 219 & -114.87 & 27.81 & 2.70 & \textbf{-0.38} & -0.43 & 0.87 & -0.29 & \textbf{-1.93} & \textbf{-1.83} \\
      \texttt{\textbf{P60468}} & 96 / 288 & -234.36 & 16.80 & 4.36 & \textbf{-0.38} & -0.02 & 0.34 & -0.14 & \textbf{-1.37} & \textbf{-2.93} \\
      \texttt{\textbf{Q9NWD9}} & 120 / 360 & -226.36 & 22.62 & 0.94 & \textbf{-0.67} & -0.01 & 0.55 & -0.07 & \textbf{-0.97} & \textbf{-0.65} \\
      \texttt{\textbf{P14555}} & 144 / 432 & -275.06 & 18.99 & 3.63 & \textbf{-0.41} & -0.09 & 0.65 & \textbf{-0.41} & \textbf{-0.29} & \textbf{-3.11} \\
      \texttt{\textbf{Q8N111}} & 149 / 447 & -335.89 & 21.86 & 1.55 & \textbf{-0.52} & -0.36 & 1.19 & -0.34 & \textbf{-2.85} & \textbf{-1.42} \\
      \texttt{\textbf{P63125}} & 156 / 468 & -299.18 & 22.01 & 2.71 & -0.40 & -0.30 & 0.13 & \textbf{-0.55} & \textbf{-2.32} & \textbf{-2.01} \\
      \texttt{\textbf{Q6XD76}} & 172 / 516 & -427.19 & 15.94 & 2.42 & \textbf{-0.63} & -0.01 & 0.57 & \textbf{-0.63} & \textbf{-1.25} & \textbf{-2.11} \\
      \texttt{\textbf{P0DMU9}} & 189 / 567 & -361.48 & 21.83 & 2.53 & \textbf{-0.64} & -0.08 & 0.66 & \textbf{-0.64} & \textbf{-1.05} & \textbf{-1.98} \\
      \texttt{\textbf{P0DPF6}} & 209 / 627 & -545.55 & 14.73 & 2.25 & \textbf{-0.51} & -0.08 & 0.14 & -0.46 & \textbf{-1.98} & \textbf{-1.93} \\
      \texttt{\textbf{Q9HD15}} & 224 / 672 & -532.98 & 15.02 & 2.26 & \textbf{-0.32} & -0.06 & 0.71 & \textbf{-0.32} & \textbf{-0.57} & \textbf{-1.98} \\
      \texttt{\textbf{Q6T310}} & 242 / 726 & -504.29 & 16.74 & 3.40 & \textbf{-0.53} & 0.01 & 0.54 & -0.27 & \textbf{-1.45} & \textbf{-2.49} \\
      \texttt{\textbf{Q9BRP0}} & 275 / 825 & -586.50 & 17.42 & 2.41 & \textbf{-0.98} & -0.51 & 0.51 & -0.55 & \textbf{-1.86} & \textbf{-1.68} \\
      \texttt{\textbf{P56178}} & 289 / 867 & -606.58 & 19.00 & 3.31 & \textbf{-0.92} & -0.03 & 0.79 & -0.91 & \textbf{-1.02} & \textbf{-2.75} \\
      \texttt{\textbf{Q8NH87}} & 305 / 915 & -572.37 & 17.15 & 7.03 & \textbf{-1.06} & -0.22 & 0.54 & -0.85 & \textbf{-1.19} & \textbf{-3.94} \\
      \texttt{\textbf{Q8NGU1}} & 314 / 942 & -612.55 & 16.58 & 6.77 & \textbf{-0.87} & -0.44 & 1.04 & -0.35 & \textbf{-0.81} & \textbf{-3.14} \\
      \texttt{\textbf{Q8NGC9}} & 324 / 972 & -582.39 & 21.08 & 9.93 & \textbf{-1.04} & -0.46 & 0.59 & -0.09 & \textbf{-1.11} & \textbf{-5.69} \\
      \texttt{\textbf{Q99729}} & 332 / 996 & -667.21 & 22.79 & 2.02 & \textbf{-1.00} & -0.29 & 0.48 & -0.54 & \textbf{-1.56} & \textbf{-1.26} \\
      \texttt{\textbf{Q9P2M1}} & 347 / 1041 & -663.54 & 22.08 & 3.83 & \textbf{-1.53} & 0.06 & 0.81 & -1.44 & \textbf{-1.93} & \textbf{-2.56} \\
      
      % ... repeat all 20 UniProt targets ...
      \addlinespace[0.8mm]
      \midrule

      % ============================
      % Spike block
      % ============================
      \multicolumn{11}{c}{\textbf{SARS-CoV-2 spike Protein}} \\
      \cmidrule(lr){1-11}
      \addlinespace[0.6mm]

      \multicolumn{2}{c}{} &
      \multicolumn{3}{c}{\textbf{\lineardesign (LD)}} &
      \multicolumn{3}{c}{\textbf{\ensembledesign (vs. LD)}} &
      \multicolumn{3}{c}{\textbf{Ours (vs. LD)}} \\
      \cmidrule(lr){1-2}
      \cmidrule(lr){3-5}
      \cmidrule(lr){6-8}
      \cmidrule(lr){9-11}

      \textbf{ID} &
      \textbf{length (aa/nt)} &
      $\Delta G^{\circ}_{\mathrm{ens}}$ &
      \AUP &
      \AccessU &
      $\Delta\Delta G^{\circ}_{\mathrm{ens}}$ &
      $\Delta\AUP$ &
      $\Delta\AccessU$ &
      $\Delta\Delta G^{\circ}_{\mathrm{ens}}\objmark$ &
      $\Delta\AUP\objmark$ &
      $\Delta\AccessU\objmark$ \\
      \midrule
      
      \texttt{\textbf{SPIKE}} & 1273 / 3819 & -2511.49 & 17.43 & 3.84 & \textbf{-4.35} & -0.21 & 0.73 & -2.47 & \textbf{-1.26} & \textbf{-2.43} \\
      
      \bottomrule
    \end{tabular}
    }% resizebox

  \label{tab:single}
  \end{minipage}
}; % node(T)

\end{tikzpicture}

\caption{\textbf{Single-Metric Optimization on UniProt Proteins and SARS-CoV-2 spike.}
\textit{(a--c)} Metric change relative to \lineardesign (baseline $=0$) versus protein length:
(a) $\Delta\Delta G^{\circ}_{\mathrm{ens}}$ (\EFE), (b) $\Delta\AUP$, and (c) $\Delta\AccessU$.
In each subfigure: red baseline: \lineardesign; Orange curve: \ensembledesign--\lineardesign; green curve: Ours--\lineardesign.
Table~\ref{tab:single} Detailed breakdown: \lineardesign reports metric values of its sequence; \ensembledesign and Ours report differences relative to \lineardesign.}

\label{fig:single}
\end{figure*}

\subsection{Protein Targets}
We evaluate on two groups of target proteins.
The first group contains $20$ natural proteins selected from UniProt~\citep{uniprot2022},
with lengths ranging from approximately $50$ to $350$ amino acids. The UniProt identifiers and lengths of the $20$ targets are listed in
Table~\ref{tab:single}.
The second target is the SARS-CoV-2 spike protein, included as an additional long
sequence.
SARS-CoV-2 spike is a high-impact and widely studied antigen in COVID-19 vaccine development,
and is commonly used as a representative long target; structurally, the SARS-CoV-2 spike
glycoprotein is a major focus of vaccine research and a primary target of host
immune defenses~\citep{bangaru2020spike}.

\subsection{Baselines}
For each target protein $\vecp$, we compare our designed sequence against
baselines produced by \lineardesign~\citep{zhang2023lineardesign} and
\ensembledesign~\citep{dai2025ensembledesign}.
For each baseline method, we use the output sequence produced by its official
implementation as the representative baseline for that target.
All baseline sequences and our sequences are re-evaluated under the same scoring
pipeline described below to ensure a fair comparison.

\subsection{Scoring Pipeline and Decoding Settings.}
All structure-informed metrics are evaluated under a consistent thermodynamic
parameter set and decoding configuration.
Specifically, \MFE is evaluated with
\texttt{LinearFold -V -d0 -b100}, and \EFE/\AUP/\AccessU are evaluated with
\texttt{LinearPartition -V -d0 -b100}.
Here \texttt{-V} selects the ViennaRNA energy parameter set to ensure all methods
are scored under the same thermodynamic model, and \texttt{-d0} disables
dangle-energy contributions.
The beam size \texttt{-b100} applies beam pruning to accelerate evaluation.
For \CAI evaluation in SARS-CoV-2 spike experiments, we use the human codon-usage frequency
table as the reference.

\subsection{Optimization Settings.}
We optimize lattice logits $\vectheta$ using the sample--evaluate--update
procedure in Section~\ref{sec:algorithm}, with Adam updates in logit space
(Eq.~\eqref{eq:adam_25}).
We use Adam with $(\beta_1,\beta_2)=(0.9,0.999)$, and set the initial learning rate to $\stepsize=0.1$ in all experiments.
Each iteration samples a batch of $\batchsize$ sequences from the current
lattice distribution and evaluates the target objective on each sample.

\subsection{Initialization of the Sampling Distribution.}
The initial lattice distribution is constructed by mixing a
\lineardesign-based distribution with a uniform distribution.
Concretely, for each lattice state $\latticestate$, we initialize the local
transition distribution as
\begin{equation}
\transprob{\vectheta_{1}}{\cdot}{\latticestate}
=
\epsilon\;\transprob{\vectheta_{\mathrm{LD}}}{\cdot}{\latticestate}
+
(1-\epsilon)\;\transprob{\vectheta_{\mathrm{unif}}}{\cdot}{\latticestate},
\label{eq:init_mixture}
\end{equation}
where $\transprob{\vectheta_{\mathrm{LD}}}{\cdot}{\latticestate}$ corresponds to
the \lineardesign solution projected onto the lattice, and
$\transprob{\vectheta_{\mathrm{unif}}}{\cdot}{\latticestate}$ assigns equal
probability to all outgoing edges in $\outedges{\latticestate}$.
The mixing coefficient $\epsilon\in[0,1]$ controls the trade-off between
initializing near the \lineardesign solution and maintaining exploration.
Unless otherwise stated, we set $\epsilon=0.5$ in all experiments.
We then set logits $\vectheta_{1}$ such that the softmax in Eq.~\eqref{eq:softmax}
matches the mixture in Eq.~\eqref{eq:init_mixture} at every state.

\subsection{Stopping Criterion and Output Selection.}
Our method includes a patience-based early stopping rule
(Section~\ref{sec:early_stop}) that terminates optimization when the best
objective value has not improved for $\patience$ consecutive iterations.
In principle, this criterion reduces runtime by avoiding unnecessary iterations
once the optimization has converged.

In the main experiments reported in this section, however, we adopt a fixed
iteration budget $\maxiters$ for each target and objective.
This choice is motivated by the stochasticity of sampling-based optimization:
because each iteration is driven by a finite batch, the best-so-far value
can exhibit occasional minute improvements even after the optimization has
effectively stabilized, which may lead to run-to-run variability in the stopping
time under purely automatic early stopping.
To obtain comparable compute budgets across targets and baselines and to reduce
the sensitivity of termination to stochastic fluctuations, we select $\maxiters$
based on preliminary convergence curves (objective value versus iteration),
choosing a conservative cutoff after which further iterations yield negligible
improvement on average.
We still track the best-so-far sequence throughout the run and report the final
output as
\begin{equation}
\vecx^\star \;\defeq\; \argmin_{\vecx\ \text{sampled during the run}} \Obj(\vecx,\vecp),
\label{eq:output_best_so_far}
\end{equation}
i.e., the best-scoring sampled sequence under the optimized objective.

%\vspace{-0.4cm}
\section{Evaluation Results}
\label{sec:eval}
\begin{figure*}[t]
\centering
\begin{tikzpicture}

% ---------- Row: three side-by-side spike trajectory plots ----------
\node (A) [anchor=north west] at (0,0)
  {\includegraphics[width=0.333\textwidth]{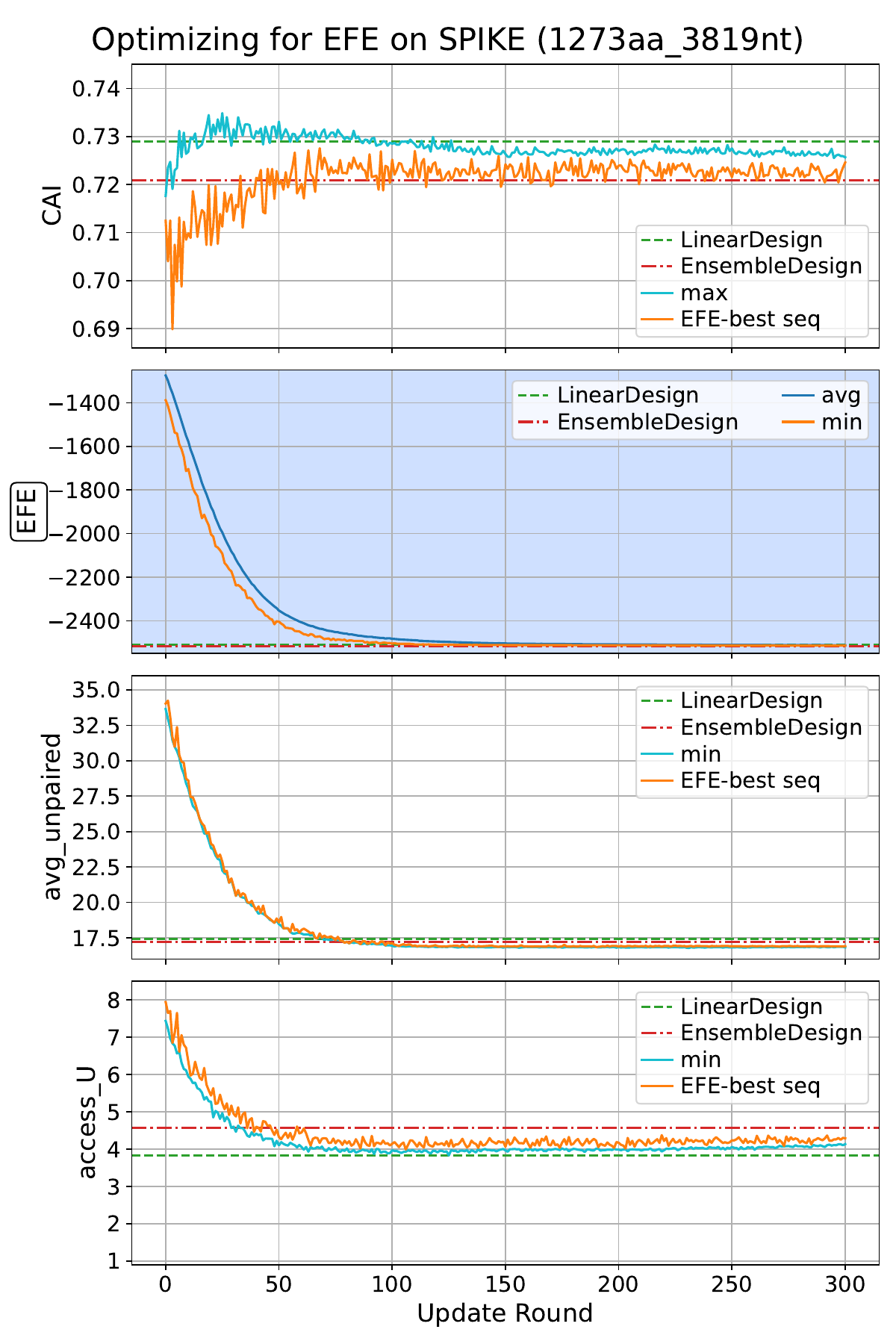}};
\node (B) [anchor=north west] at ($(A.north east)+(0,0)$)
  {\includegraphics[width=0.333\textwidth]{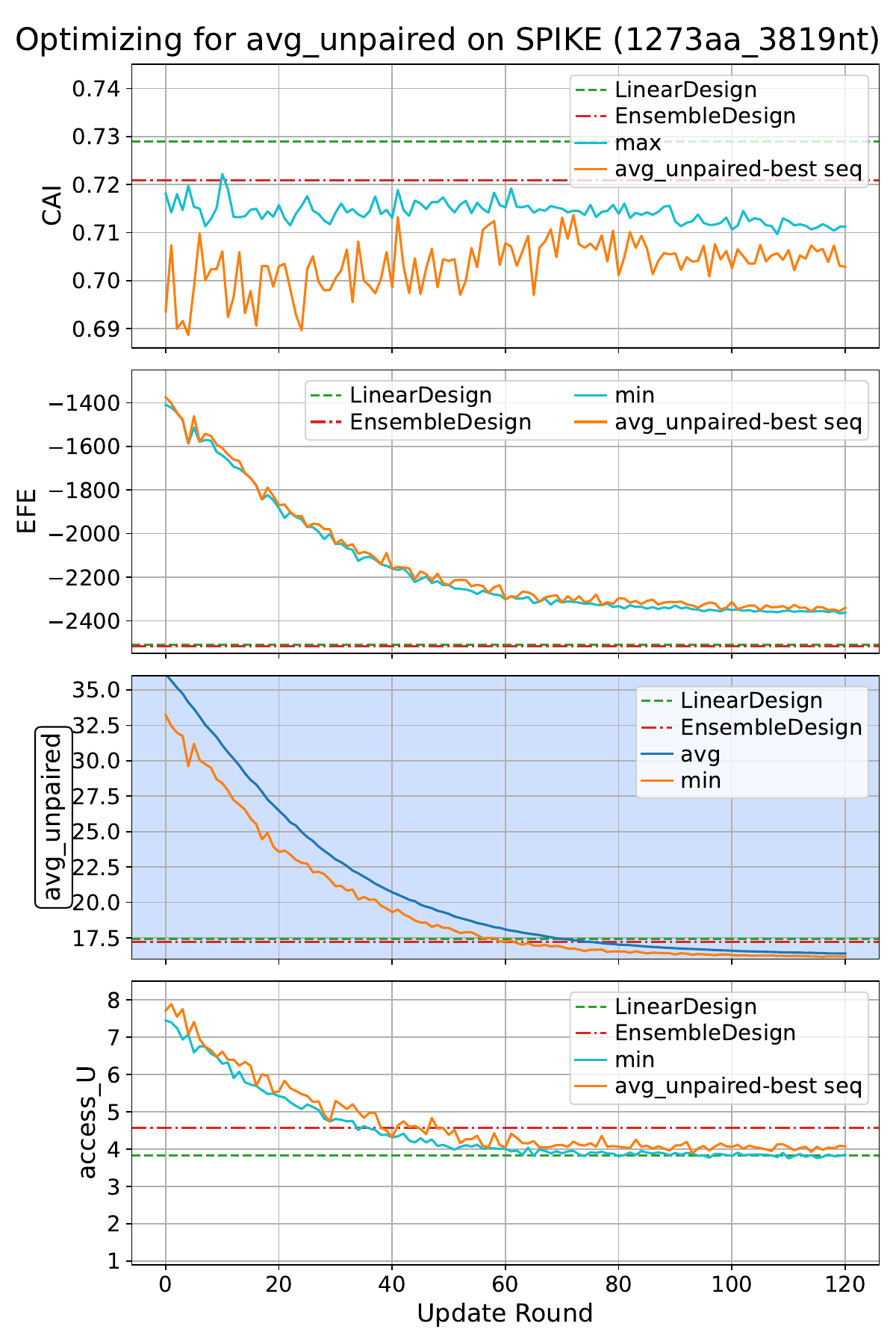}};
\node (C) [anchor=north west] at ($(B.north east)+(0,0)$)
  {\includegraphics[width=0.333\textwidth]{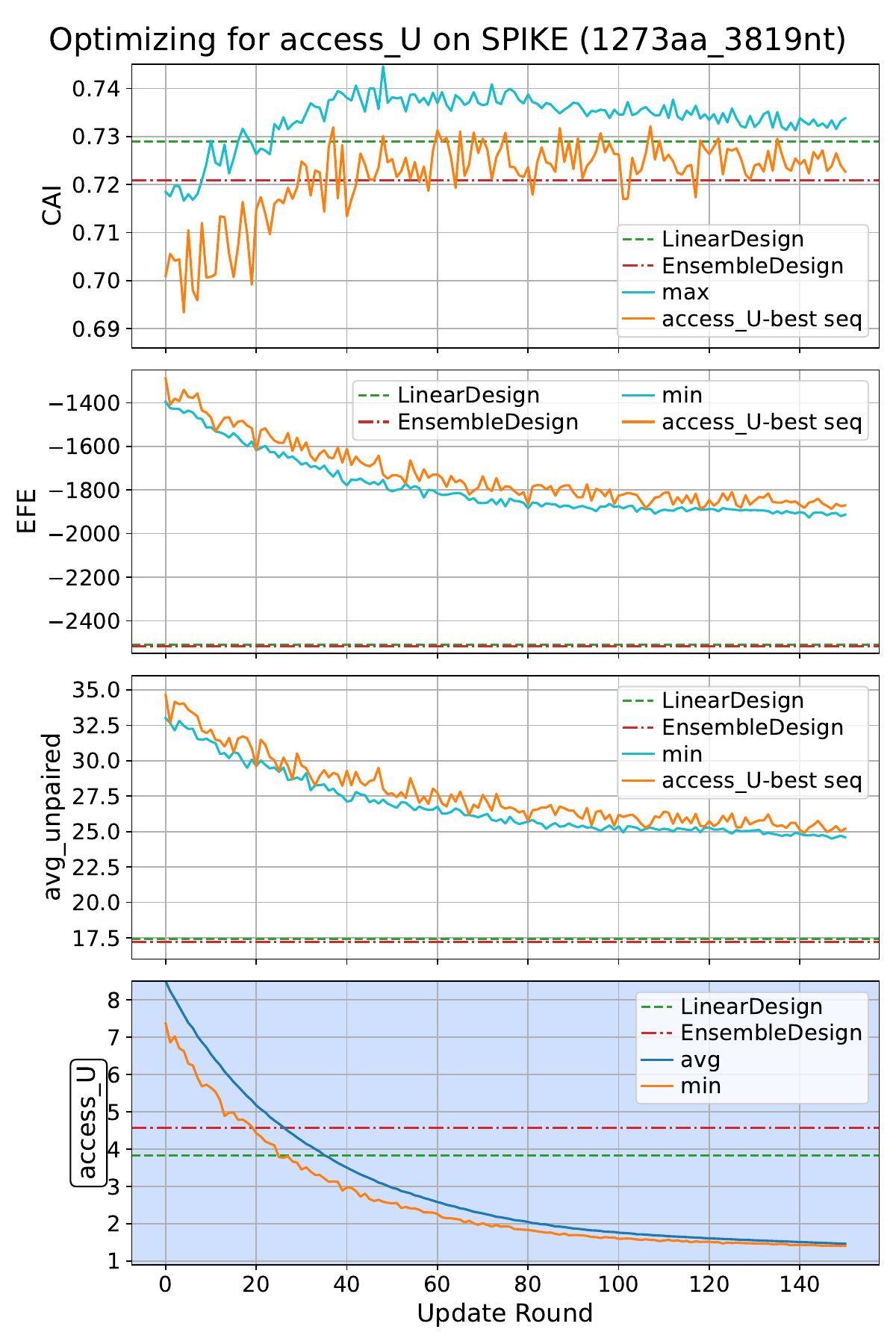}};

% ---------- Panel labels (centered, two-line to avoid overlap) ----------
\node[anchor=south, align=center] at ($(A.north)+(0,1mm)$)
  {\small (a) SPIKE \EFE optimization};
\node[anchor=south, align=center] at ($(B.north)+(0,1mm)$)
  {\small (b) SPIKE \AUP optimization};
\node[anchor=south, align=center] at ($(C.north)+(0,1mm)$)
  {\small (c) SPIKE \AccessU optimization};

\end{tikzpicture}

\caption{\textbf{SARS-CoV-2 spike trajectories for single-metric optimization.}
Panels (a)--(c) correspond to optimizing \EFE, \AUP, and \AccessU on the spike protein, respectively.
Each panel contains trajectories of sampled-sequence statistics over iterations.
Subplots with blue backgrounds indicate the primary optimized metric.
In a primary-metric subplot, the blue curve denotes the batch mean over sampled sequences, and the orange curve denotes the value of the best sampled sequence under the corresponding optimization objective.
In the remaining subplots, the orange curve reports the other metric values of that same best-by-objective sampled sequence, and the cyan curve denotes the best sampled value of that metric within each batch.}

\label{fig:single_spike_traj}
\end{figure*}

% =========================
% Figure: SPIKE COMBO comparisons in the LinearDesign design space
% =========================
\begin{figure*}[t]
  \centering
  \begin{tikzpicture}
    % ---- base image ----
    \node[inner sep=0pt] (img) {\includegraphics[width=\textwidth]{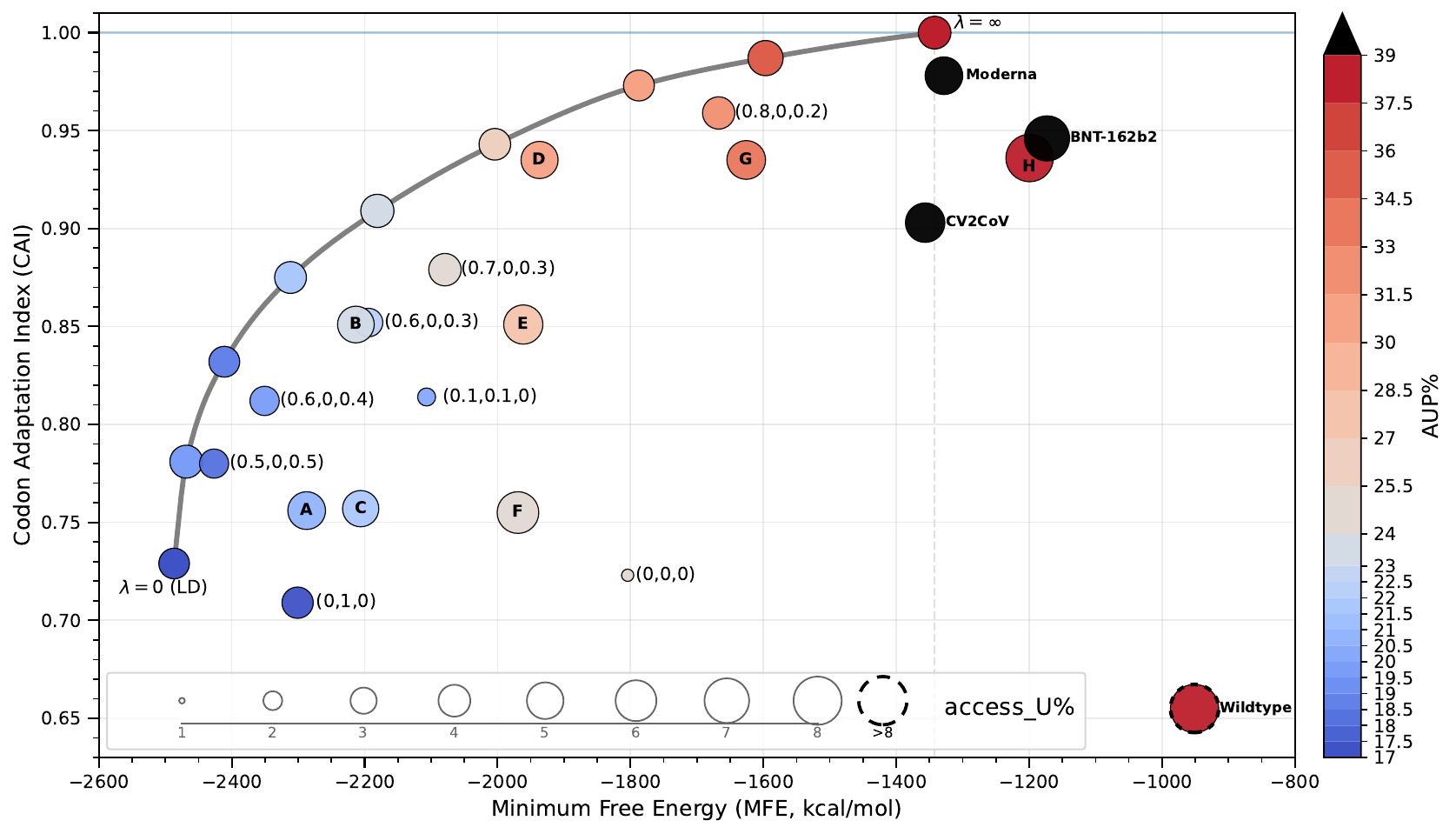}};
    
    % ===== Arrow overlay (edit the two offsets) =====
    \draw[->, line width=0.7pt]
    ($(img.north west)+(0.24\textwidth, -0.21\textwidth)$) -- ($(img.north west)+(0.2\textwidth,-0.09\textwidth)$);

    % =========================
    % Inset table 1 (top-left): Ours vs B
    % =========================
    % 你主要会调这两个数：(1) x 偏移 (2) y 偏移
    \node[anchor=north west] at ($(img.north west)+(0.055\textwidth,-0.02\textwidth)$) {%
      \fontsize{5}{6}\selectfont
      \setlength{\tabcolsep}{1.0pt}
      \renewcommand{\arraystretch}{1.00}
      \begin{tabular}{|c|c|c|c|c|}
        \hline
        & \textbf{\MFE} & \textbf{\CAI} & \textbf{\AUP} & \textbf{\AccessU} \\
        & (kcal/mol) & (\%) & (\%) & (\%) \\
        \hline
        \textbf{B} &
        \cellcolor{yellow!25}{\textbf{-2213.20}} &
        85.1 &
        23.45 &
        5.00 \\
        \hline
        \textbf{(0.6,0,0.3)} &
        -2194.10 &
        \cellcolor{yellow!25}{\textbf{85.2}} &
        \cellcolor{yellow!25}{\textbf{22.36}} &
        \cellcolor{yellow!25}{\textbf{3.36}} \\
        \hline
      \end{tabular}
    };

    \node[anchor=north west] at ($(img.north west)+(0.25\textwidth,-0.395\textwidth)$) {%
      \fontsize{5}{6}\selectfont
      \setlength{\tabcolsep}{1.0pt}
      \renewcommand{\arraystretch}{1.00}
      \begin{tabular}{|c|c|c|c|c|}
        \hline
        & \textbf{\MFE} & \textbf{\CAI} & \textbf{\AUP} & \textbf{\AccessU} \\
        & (kcal/mol) & (\%) & (\%) & (\%) \\
        \hline
        \textbf{(0,0,0)} &
        -1804.20 &
        72.3 &
        25.37 &
        \cellcolor{yellow!25}{\textbf{1.36}} \\
        \hline
        \textbf{(0,1,0)} &
        -2300.90 &
        70.9 &
        \cellcolor{yellow!25}{\textbf{16.18}} &
        3.89 \\
        \hline
      \end{tabular}
    };

  \end{tikzpicture}
  \caption{\textbf{Comparison of COMBO optimization on SARS-CoV-2 spike with prior designs in the LinearDesign design space.} The spike mRNA design space is visualized in four dimensions: minimum free energy (\MFE; x-axis) and codon adaptation index (\CAI; y-axis), with point color encoding \AUP (percent; colorbar) and circle size encoding \AccessU (percent; size legend). The gray curve is the feasibility limit (optimal boundary) computed by \lineardesign by varying the codon-optimality weight $\lambda$ from $0$ to $\infty$. Points A--G are spike sequences designed by \lineardesign as suboptimal candidates, and H is a codon-optimized baseline designed by OptimumGene. Four reference SARS-CoV-2 spike mRNA sequences are annotated for comparison: \textit{Wildtype}, \textit{BNT-162b2}, \textit{mRNA-1273 (Moderna)}, and \textit{CV2CoV}. \COMBO results are shown as points labeled by $(\alpha,\beta,\delta)$, where $(\alpha,\beta,\gamma,\delta)$ satisfy $\alpha,\beta,\gamma,\delta\ge0$ and $\alpha+\beta+\gamma+\delta=1$ (thus $\gamma=1-\alpha-\beta-\delta$).}
  \label{fig:combo_spike}
\end{figure*}

\subsection{Single-Metric Optimization}
\label{sec:single_metric}

We first evaluate our method under three single-metric objectives by optimizing
\EFE, \AUP, and \AccessU separately. 
All sequences are scored using the shared evaluation pipeline in
Section~\ref{sec:exp_setup}. 
Figure~\ref{fig:single}\textit{(a--c)} summarizes how the achieved metric values
change with protein length, reported as differences relative to the
\lineardesign baseline (thus \lineardesign appears at $0$). In each panel, the
orange curve shows \ensembledesign relative to \lineardesign, and the green
curve shows our method relative to \lineardesign.

\paragraph{Optimizing \EFE.}
Figure~\ref{fig:single}\textit{(a)} reports $\Delta\Delta G^{\circ}_{\mathrm{ens}}$
versus protein length.
We use batch size $\batchsize=500$, with
$\maxiters=50$ iterations for UniProt and $\maxiters=300$ iterations for SARS-CoV-2 spike.
Across protein targets, our method yields a consistent reduction in \EFE relative to
\lineardesign.
Compared with \ensembledesign, our \EFE values are typically slightly higher
(worse), but the gap is small on most proteins, and there exist targets where
our design matches or surpasses \ensembledesign.

\paragraph{Optimizing \AUP.}
Figure~\ref{fig:single}\textit{(b)} reports $\Delta\AUP$ versus protein length.
We use batch size $\batchsize=500$, with $\maxiters=80$ iterations for UniProt and
$\maxiters=120$ iterations for SARS-CoV-2 spike.
Here the advantage of our method is pronounced: the green curve remains
consistently below both baselines across the UniProt set and on SARS-CoV-2 spike,
indicating substantially lower \AUP.
Notably, this improvement is maintained as length increases, suggesting that the sample--evaluate--update procedure scales well for driving down global unpairedness, rather than relying on idiosyncratic gains on short sequences.

\paragraph{Optimizing \AccessU.}
Figure~\ref{fig:single}\textit{(c)} reports $\Delta\AccessU$ versus protein length.
We use batch size $\batchsize=500$, with $\maxiters=80$ iterations for UniProt and
$\maxiters=150$ iterations for SARS-CoV-2 spike.
Our method achieves consistently lower \AccessU than
both \lineardesign and \ensembledesign across all protein targets. For most targets, we achieve improvements exceeding 1\%.

\paragraph{SARS-CoV-2 spike optimization trajectories and cross-metric effects.}
In addition, Fig.~\ref{fig:single_spike_traj}(a)--(c) further illustrates the optimization dynamics on the spike protein when \EFE, \AUP, and \AccessU are optimized separately as single objectives. Our method updates the parameters of the sampling lattice, so these plots show how sequence-level statistics obtained from lattice samples evolve over iterations, rather than the trajectory of a single sequence.

Overall, all three single-objective runs improve their primary objective, while also exhibiting clear cross-metric coupling effects. When optimizing \EFE, \AUP also decreases noticeably as a byproduct, indicating that under the current folding model, shifting the lattice distribution toward lower-\EFE regions simultaneously biases sampling toward sequences with lower unpaired probability. Conversely, when optimizing \AUP, \EFE also improves to a non-negligible extent. This bidirectional behavior suggests a strong association between \EFE and \AUP in our experimental setting.

In contrast, \AccessU optimization shows a different byproduct pattern: compared with optimizing \EFE or \AUP, optimizing \AccessU leads to a more pronounced improvement in \CAI. This is reasonable because reducing \AccessU can in part be achieved by decreasing the U content of the sequence, which in turn changes the synonymous-codon usage distribution and can indirectly increase \CAI.

\subsection{\COMBO-Metric Optimization}
\label{sec:combo_metric}

Figure~\ref{fig:combo_spike} places our \COMBO-optimized SARS-CoV-2 spike designs into the
same MFE--\CAI design space used by \lineardesign~\cite{zhang2023lineardesign} , augmented with \AUP (color)
and \AccessU (circle size). By adjusting the weight tuple $(\alpha,\beta,\delta)$ (with $\gamma=1-\alpha-\beta-\delta$),
the optimization can be steered toward different trade-offs within the spike design space,
providing a practical mechanism to select sequences that prioritize different metrics. All \COMBO runs use batch size $\batchsize=200$ with the fixed $\maxiters=350$ iterations.

\paragraph{Comparison to published reference sequences.}
Relative to the wild type spike coding sequence, the \COMBO designs shown in
Fig.~\ref{fig:combo_spike} achieve simultaneous improvements across all four
displayed metrics.  Compared with widely used reference designs, including
BNT-162b2, mRNA-1273 (Moderna), CV2CoV, and the
codon-optimized baseline H, the setting $(\alpha,\beta,\delta)=(0.8,0,0.2)$
produces a competitive design that improves \MFE, \AUP, and \AccessU relative to
these references, with \CAI only slightly below mRNA-1273.

\paragraph{Weight-controlled trade-offs and examples of movement in the design space.}
Two single-metric extremes are also visible: $(0,1,0)$ emphasizes \AUP
optimization, whereas $(0,0,0)$ corresponds to emphasizing \AccessU via
$\gamma=1$. Starting from $(0,0,0)$, adding modest weight to \CAI and \AUP yields
$(0.1,0.1,0)$, improving \CAI, \MFE, and \AUP while only slightly sacrificing
\AccessU. Along a different direction, decreasing $\alpha$ while increasing
$\delta$ (with $\beta=0$ and $\gamma=1-\alpha-\delta$) induces a systematic
shift in the MFE--\CAI plane: the sequence
$(0.8,0,0.2)\!\rightarrow\!(0.7,0,0.3)\!\rightarrow\!(0.6,0,0.4)\!\rightarrow\!(0.5,0,0.5)$
moves leftward toward lower \MFE while gradually reducing \CAI, forming a smooth
trade-off curve that follows the general direction of the \lineardesign
feasibility boundary. Notably, $(0.6,0,0.3)$ lies close to
the \lineardesign sequence \textit{B} in the MFE--\CAI plane, while achieving
better \CAI, \AUP, and \AccessU with only a small trade-off in \MFE (with the
gains in \AUP and \AccessU each exceeding 1\%).

\vspace{-0.5cm}
\section{Conclusions}
\label{sec:conclusions}
%\section{Conclusion}

We introduced sampling-based continuous optimization framework for synonymous mRNA design that leverages a parameterized lattice representation of the synonymous space $\designspaceof{\vecp}$. By placing trainable logits $\vectheta$ on local lattice transitions, the method induces a distribution $\seqdist{\vectheta}(\vecx)$ supported only on feasible coding sequences, and optimizes the expected objective via an iterative sample--evaluate--update procedure.

In single-metric experiments, our method reliably drives down the targeted objective across the UniProt set and the SARS-CoV-2 spike protein under a shared scoring pipeline. Most notably, when optimizing \AUP or \AccessU, our designed sequences achieve lower values than both \lineardesign and \ensembledesign , demonstrating a consistent advantage in reducing global unpairedness and U-accessibility.

In \COMBO optimization, the explicit weight $(\alphaw,\betaw,\gammaw,\deltaw)$ enables weight-controlled navigation of the multi-objective design space: by tuning these coefficients, we can select design sequences that satisfy different optimization preferences and occupy different regions of the design space (e.g., trading off stability and codon optimality). Importantly, the framework is extensible: additional computable metrics can be incorporated into $\Obj(\vecx,\vecp)$, allowing our approach to optimize richer objective combinations beyond the current metric set.

\vspace{-0.4cm}
\bibliographystyle{natbib}
\bibliography{references}

\clearpage

\pagenumbering{roman}

\end{document}